\documentclass[fleqn,usenatbib]{mnras}

\usepackage{amsmath}	
\usepackage{amssymb}	
\usepackage{txfonts} 

\usepackage[T1]{fontenc}

\DeclareRobustCommand{\VAN}[3]{#2}
\let\VANthebibliography\thebibliography
\def\thebibliography{\DeclareRobustCommand{\VAN}[3]{##3}\VANthebibliography}

\usepackage{ulem}

\usepackage[dvipdfmx]{graphicx}	

\title[Magnetic support for CCSN explosions]{Magnetic support for neutrino-driven explosion of 3D non-rotating core-collapse supernova models}

\author[J. Matsumoto et al.]{J. Matsumoto$^{1}$\thanks{Email:jin@rk.phys.keio.ac.jp, jin@kusastro.kyoto-u.ac.jp},
Y. Asahina$^{2}$, T. Takiwaki$^{3}$, K. Kotake$^{4}$ and H. R. Takahashi$^{5}$\\
$^{1}$Keio Institute of Pure and Applied Sciences, Keio University, Yokohama 223-8522, Japan\\
$^{2}$Center for Computational Sciences, Tsukuba University, Ibaraki 305-8577, Japan\\
$^{3}$National Astronomical Observatory of Japan, Tokyo 181-8588, Japan\\
$^{4}$Department of Applied Physics and Research Institute of Stellar Explosive Phenomena, Fukuoka University, Fukuoka 814-0180, Japan\\
$^{5}$Faculty of Arts and Sciences, Department of Natural Sciences, Komazawa University, Tokyo 154-8525, Japan}

\date{Accepted XXX. Received YYY; in original form ZZZ}

\pubyear{2022}

\begin{document}
\label{firstpage}
\pagerange{\pageref{firstpage}--\pageref{lastpage}}
\maketitle

\begin{abstract}
The impact of the magnetic field on postbounce supernova dynamics of non-rotating stellar cores is studied by performing three-dimensional magnetohydrodynamics simulations with spectral neutrino transport. The explodability of strongly and weakly magnetized models of $20$ and $27$ $M_{\odot}$ pre-supernova progenitors are compared. We find that although the efficiency for the conversion of the neutrino heating into turbulent energy including magnetic fields in the gain region is not significantly different between the strong and weak field models, the amplified magnetic field due to the neutrino-driven convection on large hot bubbles just behind stalled shock results in a faster and more energetic explosion in the strongly magnetized models. In addition, by comparing the difference between the 2nd- and 5th-order spatial accuracy of the simulation in the strong field model for $27$ $M_{\odot}$ progenitor, we also find that the higher order accuracy in space is beneficial to the explosion because it enhances the growth of neutrino-driven convection in the gain region. Based on our results of core-collapse supernova simulations for the non-rotating model, a new possibility for the origin of the magnetic field of the protoneutron star (PNS) is proposed. The magnetic field is accumulated and amplified to magnetar level, that is, $\mathcal{O}(10^{14})$\,G, in the convectively stable shell near the PNS surface.
\end{abstract}



\section{Introduction}
A core-collapse supernova (CCSN) is a colossal explosion in space. It is a site where a massive star ends its life and a subsequent compact object, such as, a protoneutron star (PNS) or a black hole, is newly born. Huge gravitational binding energy ($\sim 10^{53}$\,erg) of the collapsed iron core is released during the explosion and then, roughly 1\% of the energy is transferred into the matter as the CCSN explosion energy (typically $10^{51}$\,erg $\equiv 1$\,Bethe, $1$\,B in short) through any processes. Neutrinos are expected to be the main players in this process, though other possibilities are not excluded (for example, the magnetically driven explosion).

The neutrino-heating mechanism of CCSNe, in which neutrino radiation liberated by the release of the gravitational binding energy of the stellar core heats the matter behind the stalled shock, has been extensively studied for many decades (e.g. \citealt{Colgate66}, see also \citealt{Tony12,Janka12, Kotake12, Burrows13, Foglizzo15, Mueller20} for reviews). Historically, spherically symmetric [one-dimensional (1D)] simulations for the neutrino-driven explosion of the massive stars have been failed (\citealt{Rampp00, Liebendoerfer01, Thompson03, Sumiyoshi05}, see also e.g. \citealt{Melson15a, Radice17, Mori21} for 1D successful explosions of relatively light progenitors). The great progress of CCSN simulations is the enhancement of the neutrino-heating efficiency by non-radial flows. It is almost certain that their origins are neutrino-driven/PNS convection and hydrodynamic instabilities, such as the standing accretion shock instability (SASI; \citealt{Blondin03,Foglizzo06}). Although the explosion energy reproduced by numerical simulations is systematically less than that of a typical CCSN \citep[$1$\,B,][]{Murphy19}, recent multi-dimensional self-consistent simulations have overcome the difficulty to explode massive stellar core (e.g. \citealt{Takiwaki12,Takiwaki14,Takiwaki16,Hanke13,lentz15,Bernhard15,Summa16,OConnor18b,Pan18,Ott18,Kuroda18,Nagakura19b,Nagakura20,Vartanyan19,Nakamura19,Melson20,Bollig21}).

On the one hand, the problem related to systematically small explosion energy remains in numerical modelings, but on the other hand observations of CCSNe indicate a rather broad distribution of explosion energies \citep[from $\sim 0.1$\.B to $10$\,B,][]{Muller17, Murphy19}. The diversity of CCSNe, that is, the wide range of the explosion energies, is expected to reflect the variety of properties of the progenitor, such as mass, metallicity, rotation and magnetic field. It is also a big issue to reproduce the diversity of CCSNe by numerical modelings. In order to investigate the dependence of the progenitor properties on the explosion, two-dimensional (2D) and systematic CCSN simulations of hundreds of progenitors are performed varying their mass and metallicity \citep{Nakamura15}. One key parameter to characterize the explosion is the compactness of the progenitor defined as a ratio of the mass and the enclosed radius \citep{OO11}. The higher the compactness parameter, the higher the accretion luminosity, resulting in larger diagnostic explosion energy. The rapid rotation of the progenitor determines the fate of the final evolution of the massive star. Three-dimensional (3D) and successful explosion models for rapidly-rotating progenitors did not explode in the absence of rotation \citep{Takiwaki16, Takiwaki21, Summa18}.

In addition to the compactness and rotation of the progenitor, the magnetic field is considered as a key to account for the diversity of CCSNe in other directions. Particularly, the magnetic field is expected to have a potential to generate extreme cases of CCSNe, that is, hypernovae whose kinetic energy reaches $\sim 10$\,B \citep[e.g.][]{Iwamoto98,Soderberg06} and superluminous SNe whose luminosities are $10$--$100$ times higher than those of the canonical CCSNe \citep[e.g.][]{Gal-Yam12,Nicholl13,Moriya18}. So far, these extreme events have not been explained by the neutrino heating mechanisms. Additional energy injection through magnetohydrodynamic (MHD) processes may be required to drive these events (e.g. \citealt{Wheeler02,Burrows07,Dessart08,Dessart12}).

Although the MHD explosion mechanism of the massive star has been extensively studied assuming rapid rotating precollapse cores (e.g. \citealt{Bisnovatyi-Kogan70, LeBlancWilson70, Meier76, EMuller79, Symbalisty84, Ard00, Kotake04, Sawai05, Shibata06, Sergey06, Suwa07, Takiwaki09, Takiwaki11, Obergaulinger06a, Obergaulinger06b, Winteler12,Sawai14,Sawai16, Obergaulinger17,  Obergaulinger20, Obergaulinger18, Moesta14,Moesta15,Bugli20,Kuroda20}), only several investigations of the impact of the magnetic field on the explosion are performed in the context of the slowly- and non-rotating progenitor \citep{Endeve10, Endeve12, Obergaulinger14, Muller20b, Matsumoto20, Varma22}. The validity of the slowly-rotating magnetic core of the massive star at the pre-collapse stage is supported by the stellar evolution calculations \citep{Heger05, Ott06, Langer12}. Understanding of the evolution of slowly-rotating magnetized core during the collapse is important for not only the diversity of CCSNe but also the formation of the magnetar that is a strongly magnetized NS ($B \ge 10^{14}$\,G). This is because the spin period of the observed young magnetars is relatively long ($1$--$10$\,s) and therefore, the magnetar is naturally expected to be born as a slow rotator resulting from the collapse of the magnetized core. In fact, the magnetar-class magnetic field is generated even in a slowly rotating and fully convective PNS in the recent 3D dynamo simulation (\citealt{Masada22}, see also \citealt{Raynaud20} for a rapidly rotating PNS).

In our previous work \citep{Matsumoto20}, we investigate the impact of the magnetic field on the neutrino-driven explosion in the non-rotating cores through 2D axisymmetric MHD core-collapse simulations. We extend straightforwardly it in the 3D modeling focusing on the difference in the initial magnetic field strength of the progenitor in this study. 

This paper is organized as follows: In Section~\ref{numerical methods}, we describe calculation methods and numerical setups for our 3D modeling of MHD CCSNe. In Section~\ref{results}, the explosion dynamics and signals of the neutrinos and gravitational waves (GWs) from non-rotating stellar cores are presented. We consider the origin of the magnetic field of the PNS based on our numerical results in Section~\ref{B-field of PNS}. Finally, we summarize our findings in Section~\ref{summary}.

\section{Numerical methods and models} \label{numerical methods}
\begin{figure*}
\begin{center}
\scalebox{0.6}{{\includegraphics{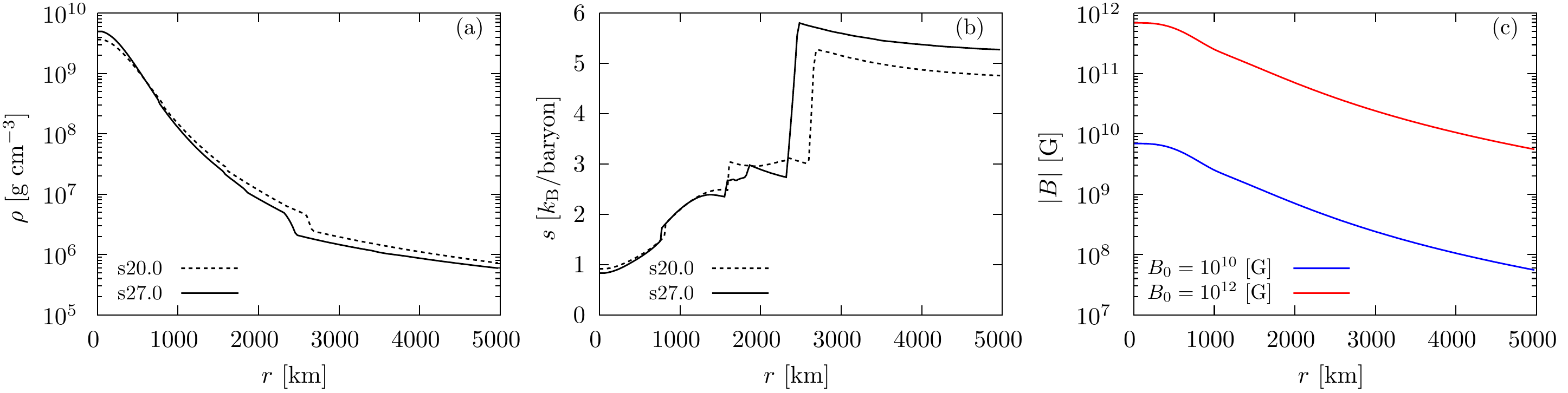}}}
\caption{The initial and radial profiles of the density (panel a), the entropy per baryon (panel b) for s20.0 (dashed line) and s27.0 (solid line) models and the spherically averaged magnetic field strength (panel c) for the strong (red) and weak (blue) field models.}
\label{fig1}
\end{center}
\end{figure*}

All 3D models in this study are calculated using our MHD supernova code, 3DnSNe (\citealt{Takiwaki16}, see \citealt{Matsumoto20} for the details of the MHD version of the code). This code, in which a three-flavour neutrino transport \citep{Kotake18} is implemented based on the isotropic diffusion source approximation (IDSA) scheme \citep{Liebendorfer09}, is designed for CCSN simulations in a 3D spherical coordinate system ($r,\theta,\phi$). The calculation methods and numerical settings in this work are essentially the same as in \citet{Matsumoto20}. We note here as the treatment for magnetofluids that the HLLD scheme \citep{Miyoshi05} is adopted to calculate numerical fluxes and the divergence cleaning method \citep{Dedner02} is implemented to reduce the numerical errors of solenoidal property of the magnetic field. In what follows, we briefly summarize some revision of the code. In order to compare the dependence of the spatial accuracy of the reconstruction of the physical variables, a 5th-order reconstruction scheme is newly implemented in our code following the numerical manner proposed in \citet{Mignone14}. In addition, the Piecewise Parabolic Method (PPM) is adapted for the slope limiter to achieve the total variation diminishing in the 5th-order reconstruction scheme. Following \citet{Mignone14}, we call it ``PPM5''. Another reconstruction scheme originally implemented in 3DnSNe is 2nd-order Piecewise Linear Method (PLM) with a modified van Leer limiter, which is named ``PLM2'' here. The 3rd and 2nd temporal accuracy is achieved in PPM5 and PLM2, respectively, using Runge--Kutta time integration.

The grid spacing and boundary conditions in this work are the same as our previous work \citep{Matsumoto20} except the number of the grid points and the grid resolution in the $\theta$- and $\phi$- direction. The calculation domain covers a sphere whose radius is $5000$\,km. The number of the grid points in the radial direction is $480$. The grid-cell size to this direction logarithmically stretches. The resolution of the polar angle is given by $\Delta({\rm cos}\theta)=$ const. with $64$ grid points covering $0 \le \theta \le \pi$. The azimuthal angle is uniformly divided into $\Delta \phi = \pi/ 64$ with $128$ grid points covering $0 \le \phi \le 2\pi$. The special treatment is imposed in the innermost $10$\,km where the spherical symmetry is assumed to avoid excessive time-step limitations. 

Two non-rotating presupernova progenitors with solar metallicity are considered in this work. One is $27.0$\,$M_\odot$ progenitor of \citet{Woosley02} that does not explode in 3D and purely hydrodynamic (HD) simulations \citep{Hanke13,Takiwaki21}, although it leads to the shock revival in both 2D HD and MHD modelings \citep{Summa16,Matsumoto20} due to the overestimation of the SASI in 2D compared to that in 3D \cite[e.g.][]{Hanke12,Hanke13,Rodrigo14}. Another is $20.0$\,$M_\odot$ progenitor of \citet{Woosley07} that is the successful model to explode fast compared to $12$--$15$ $M_\odot$ progenitors in 2D and 3D HD simulations (\citealt{OConnor18a,Burrows20}, however see \citealt{Vartanyan18} for the failed case of the explosion in 2D simulation). The radial distributions of the density and the entropy per baryon of $27.0$\,$M_\odot$ and $20.0$\,$M_\odot$ progenitors in our calculation range are plotted in Fig.~\ref{fig1} by solid and dashed lines, respectively.

Our setups for microphysics of the $27.0$ $M_\odot$ model are the same as those of \cite{Matsumoto20}. As the neutrino reaction rate, set5a of \cite{Kotake18} which is similar to that in \cite{OConnor18c} is adopted. In addition to the standard opacity set of \cite{Bruenn85}, the weak magnetism, recoil correction \citep{Horowitz02} and nucleon-nucleon bremsstrahlung are considered. The neutrino reaction rate for the $20.0$ $M_\odot$ model is almost the same as set-all of \cite{Kotake18} in which the quenching of the axial-vector coupling constant at high densities, strangeness-dependent contribution and the many-body correction to the neutrino-nucleon scattering are added to set5 to enhance the explodability. However, reactions of electron-neutrino pair annihilation into $\mu$/$\tau$ neutrinos and $\mu$/$\tau$-neutrino scattering on electron (anti)neutrinos are switched off in this work to save computational resources. This is because although these reactions does not have large impact on the neutrino luminosity or the dynamics of the fluid in our CCNS simulations, it takes roughly $1.7$ times longer to calculate 3D CCSN model after the bounce switching on such reactions in our code. Energy bins of neutrinos logarithmically spread between $1$ to $300$\,MeV. The equation of state of \cite{Lattimer91} with a nuclear incomprehensibility of $K = 220$\,MeV is used in our runs.

The initial configuration of the magnetic field in this work is the same as our previous work \cite{Matsumoto20}. The magnetic field is uniform where $r < 1000$\,km while its shape is like dipole outside the core of the massive star ($r > 1000$\,km). The explicit expression of this field is given by only a $\phi$-component of a vector potential \citep{Suwa07, Takiwaki09, Obergaulinger14, Matsumoto20} as follows;
\begin{eqnarray}
A_{\phi} = \frac{B_0}{2} \frac{r^3_0}{r^3 + r^3_0} r \sin \theta \;. \label{eq: initial vector potential}
\end{eqnarray}
Here $r_0 = 1000$\,km and we set $B_0 = 10^{12}$\,G (strong field model) or $10^{10}$\,G (weak field model) in this work. The radial distribution of the spherically averaged this field is plotted in Fig.~\ref{fig1}(c). Red and blue lines correspond to the strong and weak field models. The model name is labelled as `s27.0B10PPM5', which represents the solar metallicity $27.0$ $M_\odot$ model with $B_0=10^{10}$\,G calculated through PPM5 scheme.

In order to save computational resources, we perform 2D simulations for the core-collapse in our models until the stalled shock is excited in advance of the 3D calculations. Then, we start our 3D simulations by remapping the 2D calculation data at $10$\,ms after the bounce to 3D grids. To induce non-spherical instabilities, we add large perturbations at the timing of the remapping of the data.

\section{Results} \label{results}
There is no large difference of the progenitor dependence on our basic results, such as the dynamics of the fluid, neutrino activities and signals of the neutrinos and gravitational waves in our calculations. In particular, the tendency toward the support of the explosion by the strong magnetic field that is the main result in this study is obtained both between the $20$ and $27$ $M_\odot$ progenitor models. We choose s27B12PPM5 as a fiducial model in this work. We first present an overview of a non-rotating and strongly magnetized core-collapse model of the $27$ $M_\odot$ progenitor (s27B12PPM5) in Section~\ref{overall evolution}. Then, the mechanism of the fast explosion of the strongly magnetized model is explained in Section~\ref{explosion dinamics}. In Section~\ref{dependence of spatial accuracy}, we investigate the impact of the spatial accuracy of the simulations on the explosion. Comparing the strongly magnetized cases in $27$ $M_{\odot}$ progenitor models with different accuracy, we found that the high spatial accuracy of the simulations was positive for the explosion, that is, the fast expansion of the shock wave and large explosion energy. Finally, signals of the neutrinos and gravitational waves from $27$ $M_{\odot}$ progenitor model is calculated in Section~\ref{signals}.

\subsection{Overall evolution of 3D non-rotating and strongly magnetized core-collapse model} \label{overall evolution}

\begin{figure*}
\begin{center}

\scalebox{0.48}{{\includegraphics{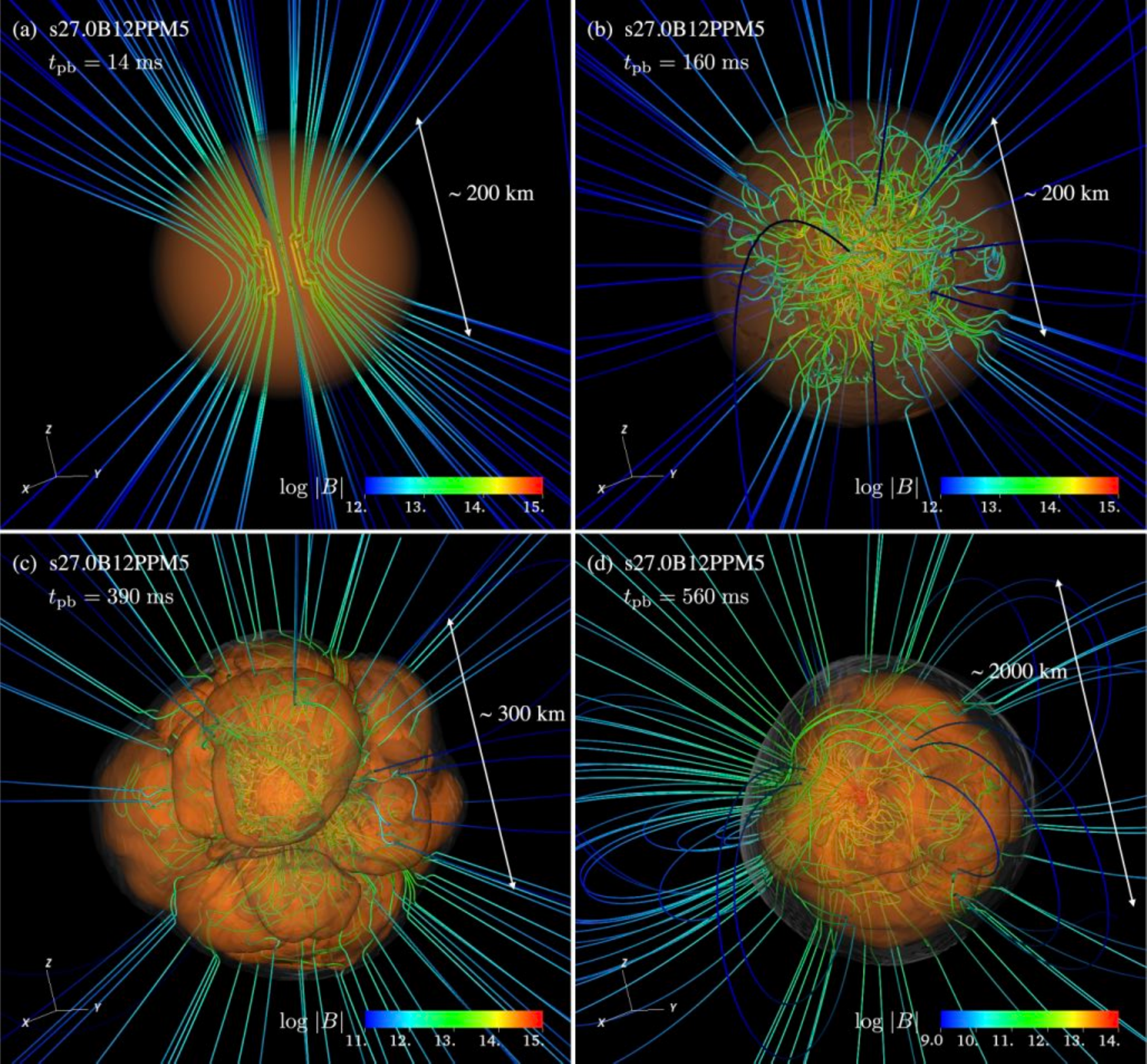}}}
\caption{Time evolution of the shock surface (outermost whitish sphere), isosurface of the entropy per baryon and magnetic field lines for fiducial model (s27.0B12PPM5). Panels (a), (b), (c) and (d) correspond to $t_{\rm pb}=14$, $160$, $390$ and $590$\,ms, respectively. Note that $t_{\rm pb}$ represents the post-bounce time. The iso-entropy surface of $10 \; k_{\rm B}$ and $15 \; k_{\rm B}$ are illustrated by brown transparent shells in panels (a), (b) and (c), (d), respectively. The viewing angle of the central object is fixed for all panels. The spatial scale is represented by the white two-headed arrow that is parallel to the $z$-axis in each panel.}
\label{fig2}
\end{center}
\end{figure*}

Fig.~\ref{fig2} shows four snapshots of the 3D evolution of a shock wave, the entropy per baryon and magnetic field lines for model s27.0B12PPM5. Panels (a), (b), (c) and (d) correspond to the time $t_{\rm pb}=14$, $160$, $390$ and $560$\,ms after bounce, respectively. Note that $t_{\rm pb}$ represents the post-bounce time. The viewing angle is fixed in all panels. The spatial scale is displayed with a white two-headed arrow that is parallel to the $z$-axis. Note that the spatial scale in panels (a) and (b) is same while that in panels (c) and (d) becomes large as time passes. An outermost whitish and transparent sphere indicates the position of the shock wave in each panel. A transparent brown shell represents the iso-entropy surface of $10 \; k_{\rm B}$ and $15 \; k_{\rm B}$ in panels (a), (b) and (c), (d), respectively. The colors of magnetic field lines denote the magnetic field strength.

\begin{figure*}
\begin{center}
\scalebox{0.9}{{\includegraphics{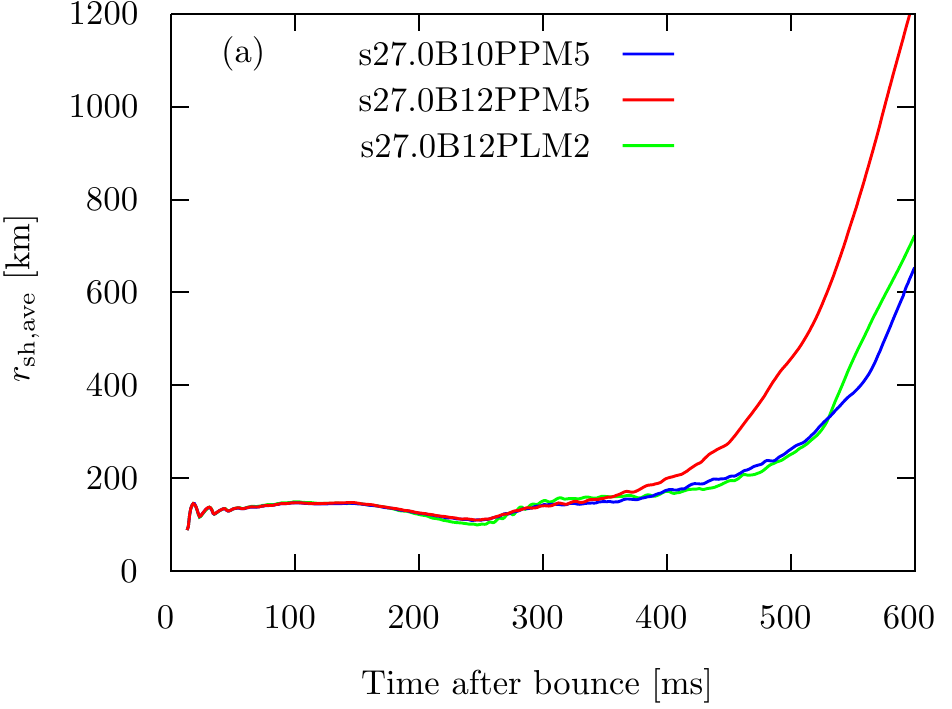}}}
\scalebox{0.9}{{\includegraphics{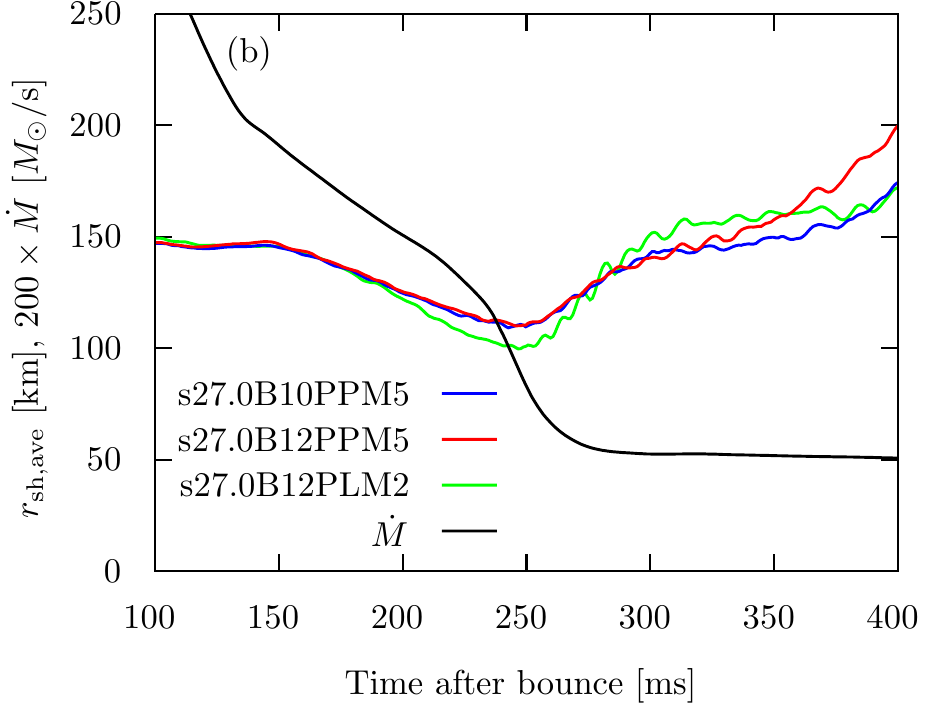}}}
\caption{Time evolution of the averaged shock radius, $r_{\rm sh,ave}$, and the mass accretion rate at $r=500$\,km, $\dot{M}$ (black line in panel b), in the  s27.0 models.}
\label{fig3}
\end{center}
\end{figure*}

Since the shock wave is excited after the bounce, the material behind it is heated and the entropy drastically grows there. The isosurface of the entropy per baryon in panels (a) and (b) is located at just behind the shock wave. The averaged shock radii in panels (a), (b), (c) and (d) are $120$, $140$, $190$ and $850$\,km, respectively (see also, Fig~\ref{fig3}a).

\begin{figure*}
\begin{center}
\scalebox{0.9}{{\includegraphics{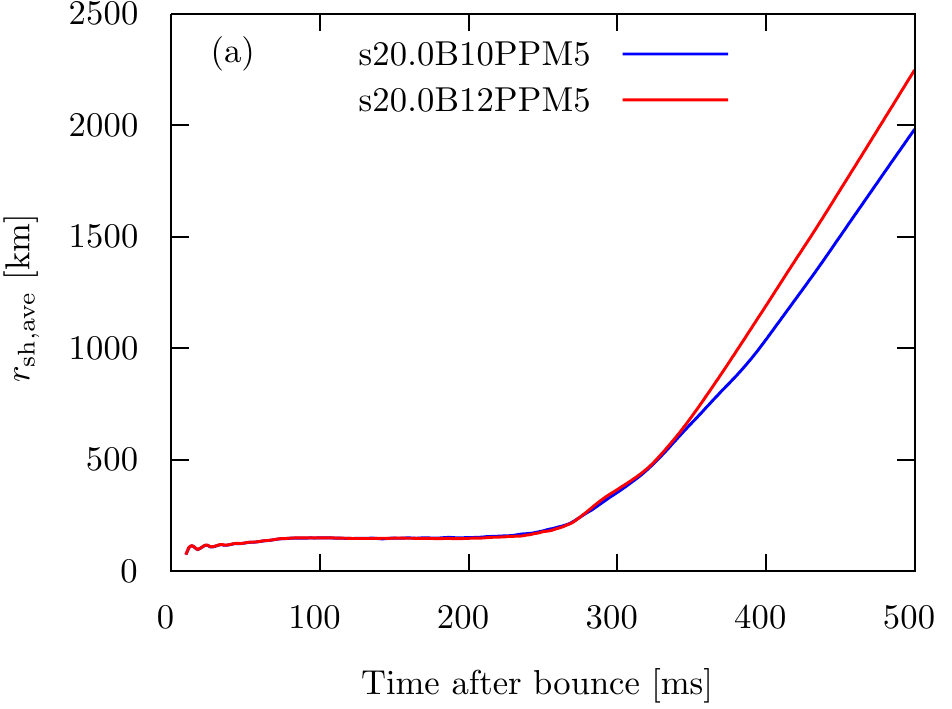}}}
\scalebox{0.9}{{\includegraphics{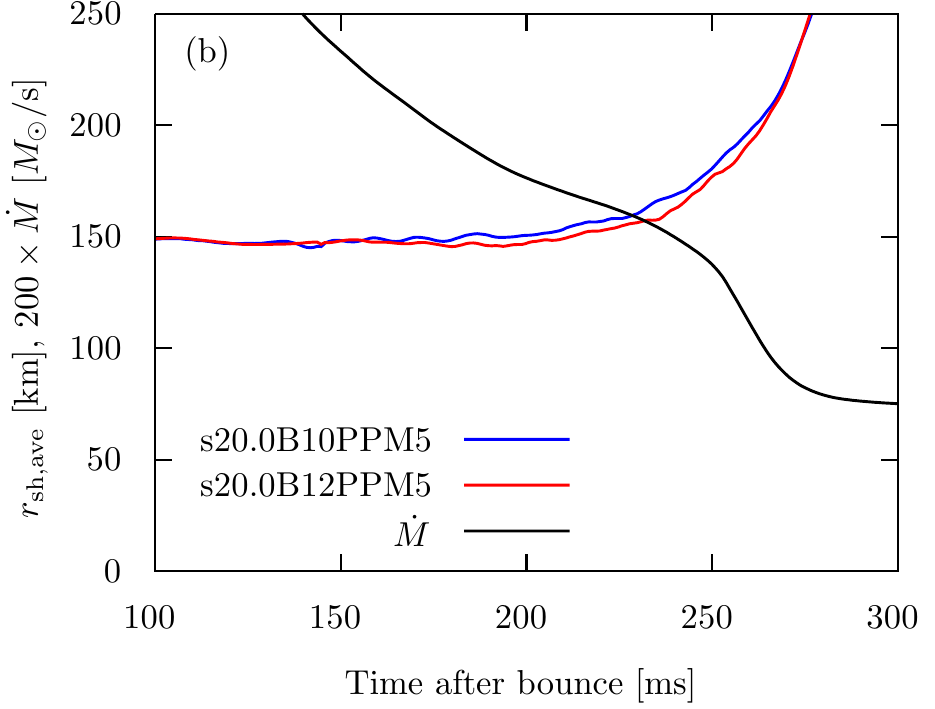}}}
\caption{Time evolution of the averaged shock radius, $r_{\rm sh,ave}$, and the mass accretion rate at $r=500$\,km, $\dot{M}$ (black line in panel b), in s20.0 models.}
\label{fig4}
\end{center}
\end{figure*}

Except for the central region of the massive star, the split-monopole-like configuration of the magnetic filed is formed in the early phase of the simulation, which corresponds to (a) in Fig.~\ref{fig2}. This is because the magnetic field is frozen-in to the fluid whose motion is restricted to the radial direction due to the gravitational collapse of the core of the massive star. The electric resistivity of the magnetic field is ignored in this study because it is expected to be small in the massive stellar core \citep{Sawai13r}. The magnetic field strength is simply amplified due to the magnetic flux conservation during the early phase of the core collapse and exceeds $10^{14}$ G within $r=30$ km.

The bounce shock stalls at $r \sim 150$\,km around $t_{\rm pb}=100$\,ms and then gradually shrinks. As shown in Fig.~\ref{fig2}(b) ($t_{\rm pb}=160$\,ms), in this phase, the structure of the magnetic field lines is still like a split monopole outside the shock surface while it is rather complicated inside the shock. This is because the magnetic field lines are bent due to the neutrino-driven and PNS convection, which originates from the negative gradient of the entropy and the lepton fraction, respectively. Therefore, the non-radial component of the magnetic field is generated. In addition, the field amplification occurs because the the magnetic field lines are compressed and stretched by the convective motion of the matter.

The panel (c) in Fig.~\ref{fig2} is the snapshot after the shock revival ($t_{\rm pb} =390$\,ms). The radius of the shock wave clearly expands compared to that in panels (a) and (b) before the shock revival. The structure of the isosurface of the entropy per baryon is not a simple sphere but more complicated. The large scale convective motion, whose origin is the neutrino heating, drastically rolls up the material from inside the core to beneath the shock radius. This leads to the large hot bubble structure of the 3D iso-entropy surface. The magnetic field lines penetrated inside the shock surface from the upper stream are along the isosurface of the entropy. They trace the trajectory of the fluid motion that forms the large bubble. The magnetic field lines are accumulated around the down flow region between bubbles. In the deep inside of the shock wave, the small scale structure of the magnetic field lines due to the PNS convection is seen.

The 3D structure of the explosion of the stellar core in the late phase of the fiducial run (s27.0B12PPM5) is shown in Fig.~\ref{fig2}(d). The shock wave finally reaches at a radius of $\sim 1000$\,km. Its shape is almost a sphere. It implies that the neutrino-driven convection mainly contributes to the enhancement of the neutrino-heating efficiency compared to the SASI that indicates an aspherical shock surface and any special direction depending on the most growing mode. The magnetic loops observed around the equatorial region (the equatorial plane is defined as $x$-$y$ plane at $z=0$) are the remnant of the initial magnetic field configuration given by the vector potential (equation \ref{eq: initial vector potential}) that forms magnetic loops on the equatorial region at large scale ($r \ge 1000$\,km).

\subsection{Fast explosion of strongly magnetized model} \label{explosion dinamics}
In this section, the mechanism of the fast explosion of the strongly magnetized core is explained focusing on the fiducial progenitor model (s27.0) because the significant difference of the progenitor dependence on this mechanism is not observed in this work. The comparison of models with different spatial accuracy is mainly performed in Section~\ref{dependence of spatial accuracy}.

Fig.~\ref{fig3}(a) shows the temporal evolution of the averaged shock radius in the $27$\,$M_\odot$ progenitor models. The blue, red and green lines represent the case for models s27.0B10PPM5, s27.0B12PPM5 and s27.0B12PLM2, respectively. In all models, the outward propagating shock wave stalls at $r \sim 150$\,km around $t_{\rm pb}=100$\,ms. Then it shrinks until $t_{\rm pb}=250$\,ms after the standing shock phase for $50$\,ms. The shock revival occurs at around $t_{\rm pb}=250$\,ms and the shock wave starts to propagate outward again.

After the shock revival, the evolution of the averaged shock radius is different between models. Compering to the dependence of the initial magnetic field strength, the shock evolution in the strong ($B_0=10^{12}$ G) field model (red line) is faster than that in the weak ($B_0=10^{10}$ G) field model (blue line). 

\begin{figure}
\begin{center}
\scalebox{0.88}{{\includegraphics{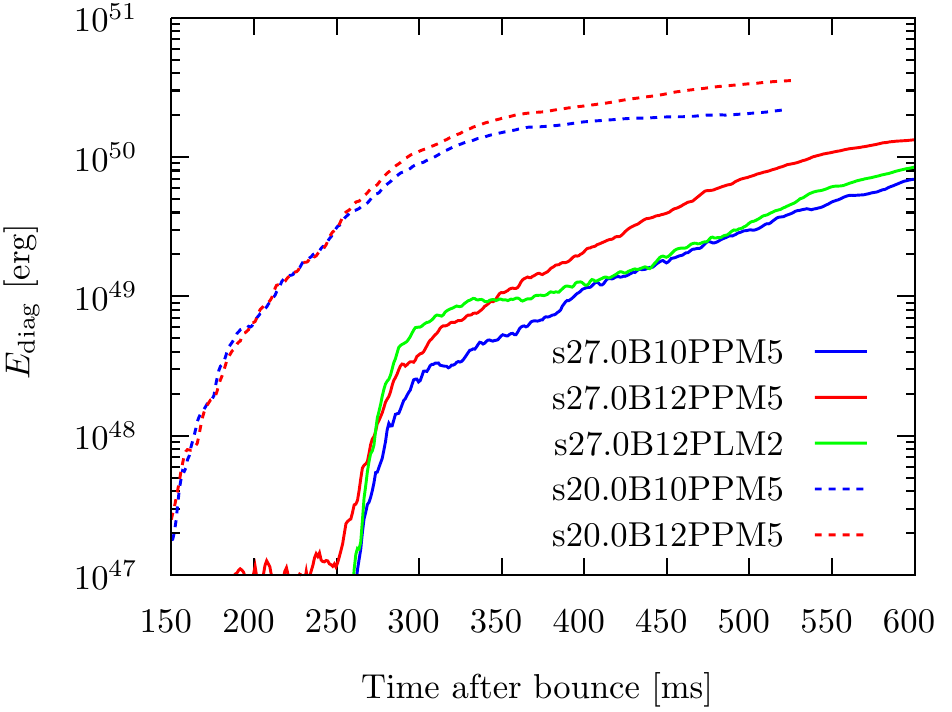}}}
\caption{Comparison of the diagnostic explosion energy defined by equation (\ref{diagnostic explosion energy}) in all our runs. Solid and dashed lines correspond to the $27$ and $20$ $M_{\odot}$ progenitor models, respectively. Red and blue lines are the cases for the strong and weak field models, respectively, in each progenitor model.}
\label{fig5}
\end{center}
\end{figure}

To focus on the difference of models around the onset of the shock revival, the shock evolution between $t_{\rm pb}=100$\,ms and $t_{\rm pb}=400$\,ms is shown in Fig.~\ref{fig3}(b). The colors have the same meaning as Fig.~\ref{fig3}(a). In addition to the time evolution of the shock radius, the temporal evolution of the mass accretion rate at $r=500$\,km, $\dot{M}$, is represented by a black line that is almost the same in all models. At around $t_{\rm pb}=250$\,ms, $\dot{M}$ drastically decreases. This indicates that a Si/O-rich layer in which the density is much less than that in the Fe core collapses to the inside of the shock. Since the ram pressure for the shock surface in the upper stream also decreases, it is reasonable that the shock revival occurs around the sudden drop in the mass accretion rate. This is consistent with the shock evolution for $27$\,$M_\odot$ model of \citet{Hanke13}. Here, we stress that the shock expansion in the 3D HD model of \citet{Hanke13} continues only for $30$\,ms after the drastic decrease in the mass accretion rate and turns into a contraction while the shock wave in the 2D HD model successfully expands.

\begin{figure}
\begin{center}
\scalebox{0.89}{{\includegraphics{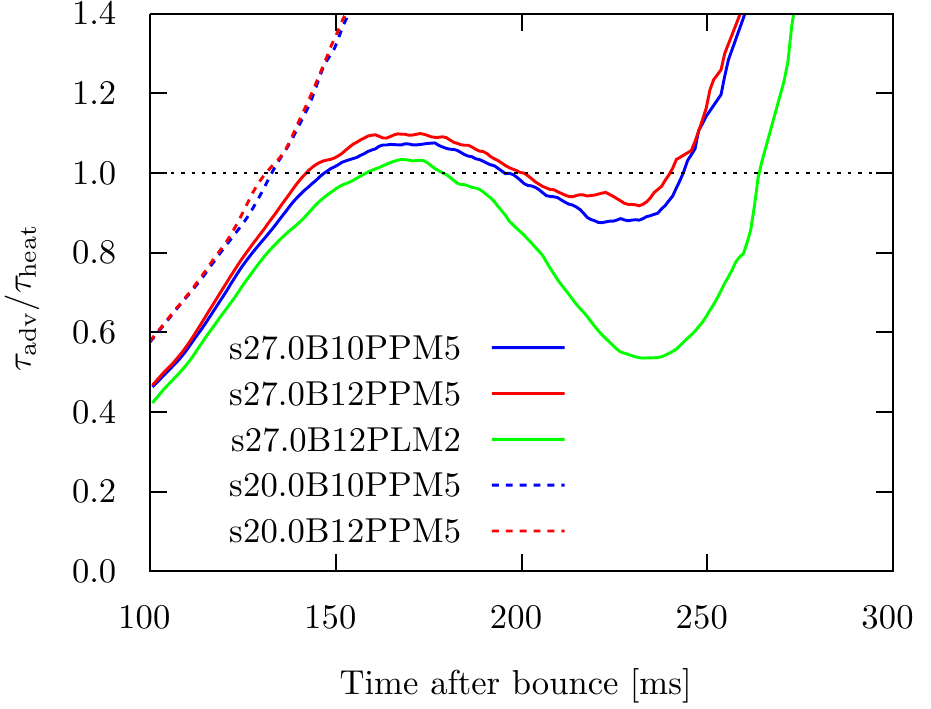}}}
\caption{Time evolution of the ratio of advection timescale to heating timescale in all models at around the onset of the shock revival. Line types and colours have the same meanings as those in Fig.~\ref{fig5}}.
\label{fig6}
\end{center}
\end{figure}

\begin{figure*}
\begin{center}
\scalebox{0.9}{{\includegraphics{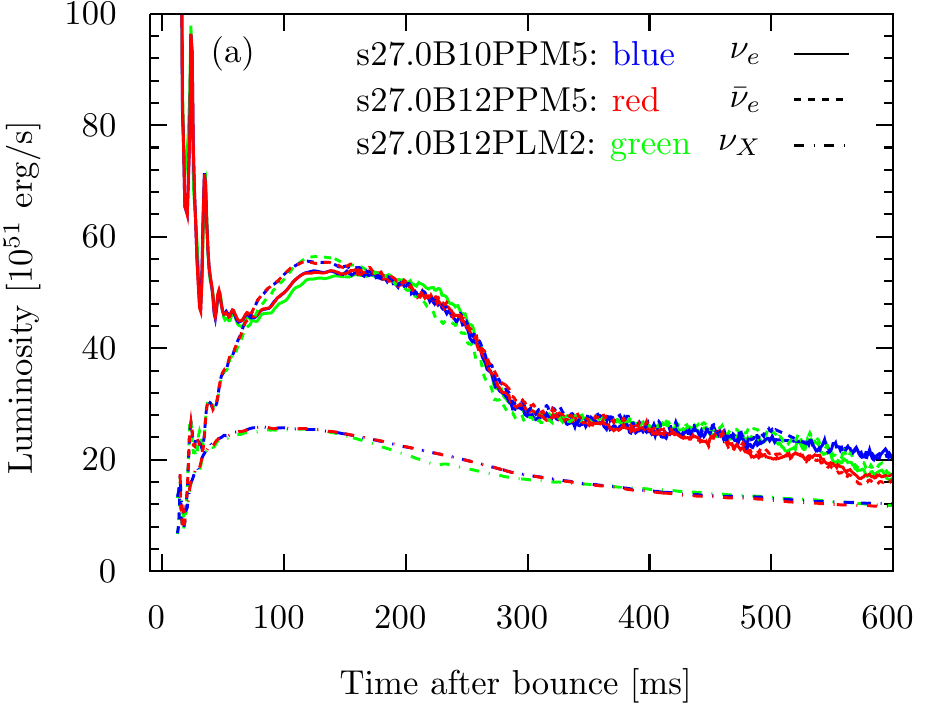}}}
\scalebox{0.9}{{\includegraphics{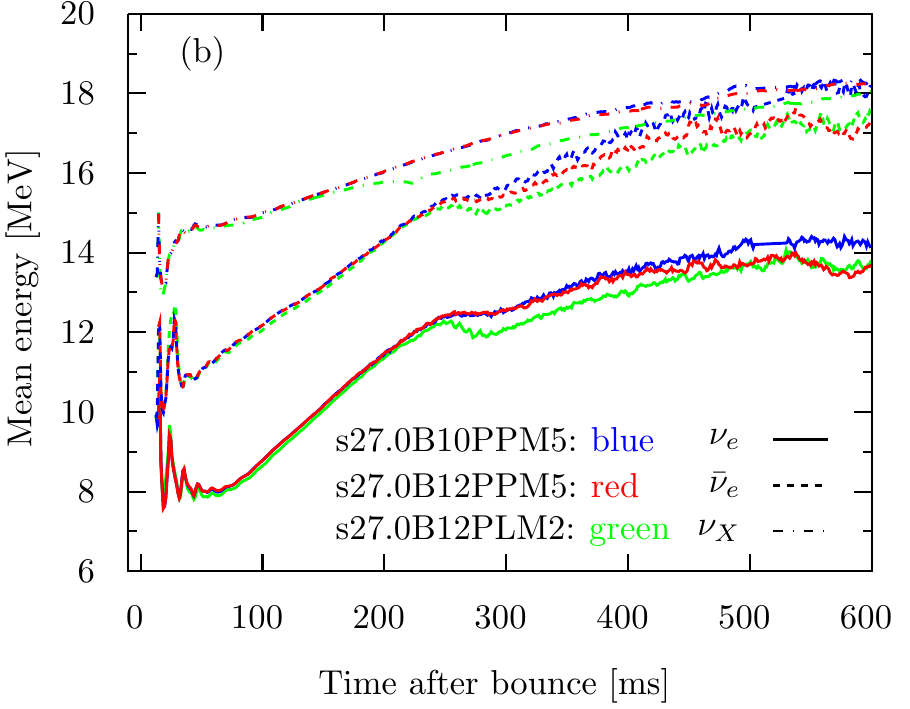}}}
\caption{Time evolution of (a) neutrino luminosity and (b) mean energy in the fiducial progenitor model (s27.0).}
\label{fig7}
\end{center}
\end{figure*}

\begin{figure}
\begin{center}
\scalebox{0.88}{{\includegraphics{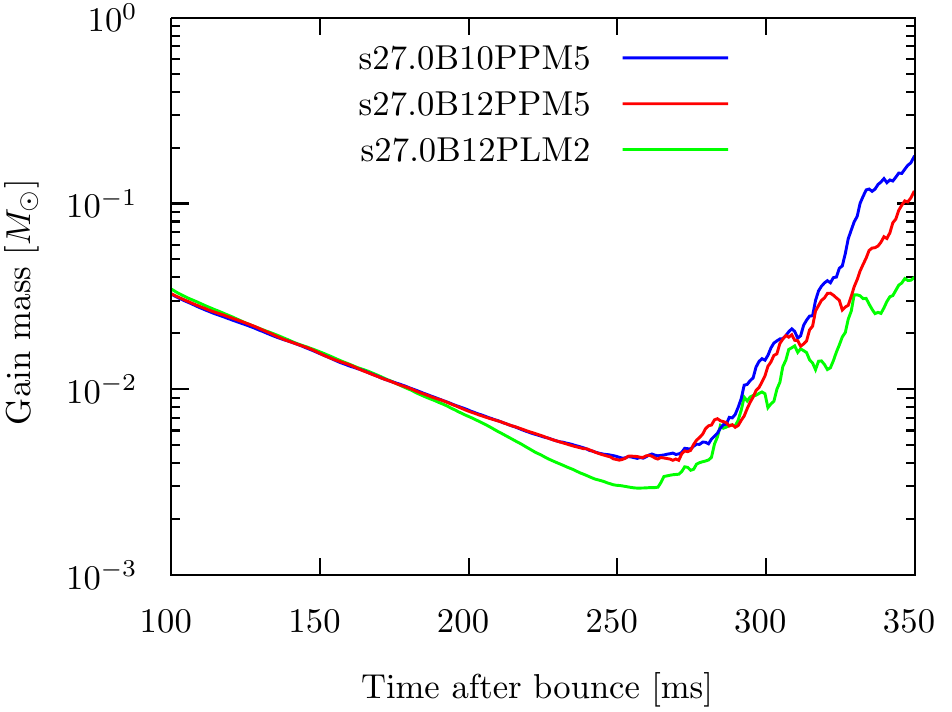}}}
\caption{Time evolution of the gain mass in $27$ $M_{\odot}$ progenitor models at around the shock revival.}
\label{fig8}
\end{center}
\end{figure}

Fig.~\ref{fig4} shows the temporal evolution of the averaged shock radius in 20 solar mass progenitor models. In panel (b), the evolution only close to the shock revival time (between $t_{\rm pb}=100$\,ms and $t_{\rm pb}=300$\,ms) is shown and the time evolution of the mass accretion rate at $r=500$\,km is also depicted. The blue and red lines indicate the shock evolution of models s20.0B10PPM5 and s20.0B12PPM5, respectively. The black line corresponds to the mass accretion rate. In s20.0 series models, the rapid expansion of the shock surface is also observed when the mass accretion rate drastically drops due to the gravitational collapse of the Si/O layer to the inside of the shock surface. The dependence of the initial magnetic field of the progenitor on the evolution of the shock surface after the shock revival is the same as the series of $27$ $M_{\odot}$ progenitor models. The shock surface expands fast in the strong ($B_0=10^{12}$\,G) field model (red line) compared to that in the weak ($B_0=10^{10}$\,G) field model (blue line). 

The explosion energy in the fast explosion model is expected to becomes large because the mass accretion rate is high in the early phase of the core collapse. Actually, it decreases as time passes as shown in Fig.~\ref{fig3}(b) and Fig.~\ref{fig4}(b). Therefore, the neutrino luminosity that is the source of the explosion energy would be also large in the fast explosion model. The temporal evolution of the diagnostic explosion energy of $27$ $M_{\odot}$ (solid lines) and $20$ $M_{\odot}$ (dashed lines) progenitor models are shown in Fig.~\ref{fig5}. The meanings of the colours for lines of s27.0 and s20.0 models are the same as those in Fig.~\ref{fig3} and Fig.~\ref{fig4}, respectively. The red and blue lines represent the strong and weak field models, respectively, in each progenitor model. The diagnostic explosion energy is defined by
\begin{eqnarray}
E_{\rm diag} = \int_{D} \biggl [ \rho \biggl ( \frac{1}{2}v^2 + \epsilon_{\rm int} + \Phi \biggr ) + \epsilon_{\rm mag} \biggr ] {\rm d} V \; , \label{diagnostic explosion energy}
\end{eqnarray}
where $\rho$ is the mass density, $v$ is the velocity, $\epsilon_{\rm int}$ is the specific internal energy, $\Phi$ is the gravitational potential, $\epsilon_{\rm mag}$ is the magnetic energy density and $V$ is the volume of the calculation cell, respectively. The integration is performed in the region $D$ where the integrand is positive. Note that the integrand in equation (\ref{diagnostic explosion energy}) includes the magnetic energy density in addition to the ordinary definition of it \citep[see e.g.][]{Suwa10,Takiwaki21}. However, the contribution of the magnetic energy to the diagnostic explosion energy is extremely small. This is because the thermal energy is dominant in our system. As expected, the diagnostic explosion energy of the strong field models (red lines) is greater than that of the weak field models (blue lines) in each progenitor models. 

To quantify the neutrino-heating efficiency, the temporal evolution of the timescale ratio between advection and heating in $20$ and $27$ $M_{\odot}$ progenitor models around the shock revival phase is shown in Fig.~\ref{fig6}. The line types and colours have the same meanings as those in Fig.~\ref{fig5}. Following \citet{Summa16} and \citet{Matsumoto20}, we define the advection timescale as
\begin{eqnarray}
\tau_{\rm adv} = \frac{M_{\rm g}}{\dot{M}} \; ,
\end{eqnarray}
where $M_{\rm g}$ is the gain mass. The neutrino-heating timescale is estimated by
\begin{eqnarray}
\tau_{\rm heat} = \frac{|E_{\rm tot,g}|}{\dot{Q}_{\rm heat}} \; ,
\end{eqnarray}
where $|E_{\rm tot, g}|$ and $\dot{Q}_{\rm heat}$ are the total energy of the material and the neutrino-heating rate, respectively, in the gain layer.

The ratio of the advection timescale to the heating timescale, $\tau_{\rm adv}/\tau_{\rm heat}$, is considered to be a good indicator for the neutrino-driven explosion \citep[e.g.][]{Buras06}. This is because the residency time of the matter in the gain region relatively becomes longer when the advection timescale is longer than the heating timescale. This means that the matter is effectively exposed by neutrinos and can gain the substantial thermal energy to overcome the ram pressure due to the mass accretion in the upper stream of the shock surface.

The timescale ratio excessively exceeds unity after $t_{\rm pb}=250$\,ms in the model s27.0 series. This links to the shock evolution, that is, the onset of the shock revival at around $t_{\rm pb} \sim 250$\,ms (see Fig.~\ref{fig3}b). In the model s20.0 series, the timescale ratio rapidly grows beyond unity at around $t_{\rm pb} \sim 130$\,ms. Responding to that, the shock surface gradually starts to propagate outward after $t_{\rm pb} \sim 130$\,ms while it drastically expands at the timing of the sudden drop of the mass accretion rate as shown in Fig.~\ref{fig4}(b). The association between the evolution of the timescale ratio and shock radius indicates that the neutrino-driven explosion occurs in all our models.

\begin{figure*}
\begin{center}
\scalebox{0.9}{{\includegraphics{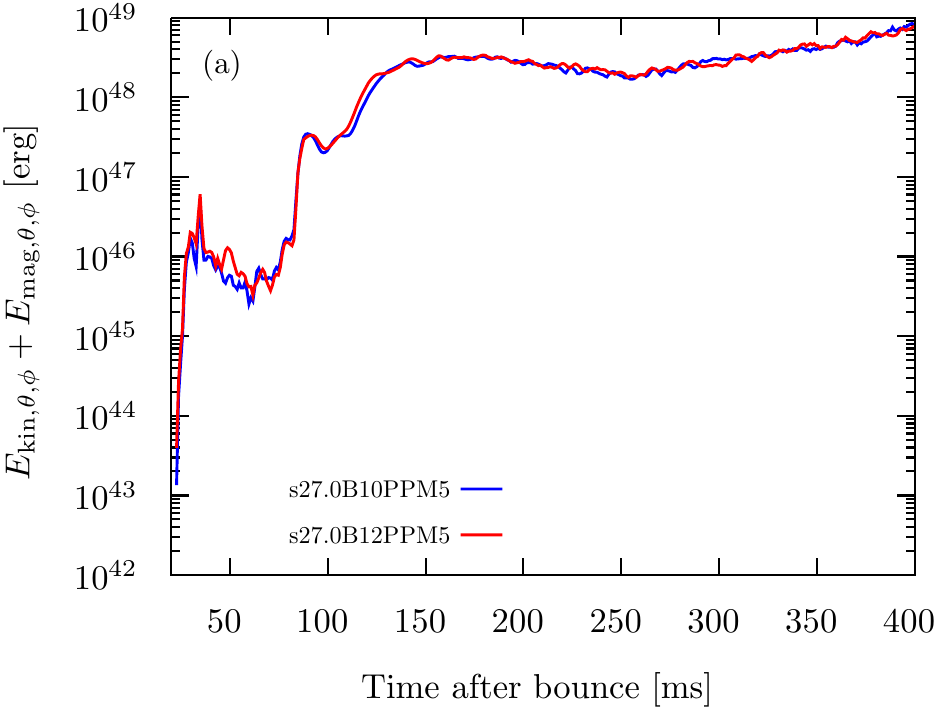}}}
\scalebox{0.9}{{\includegraphics{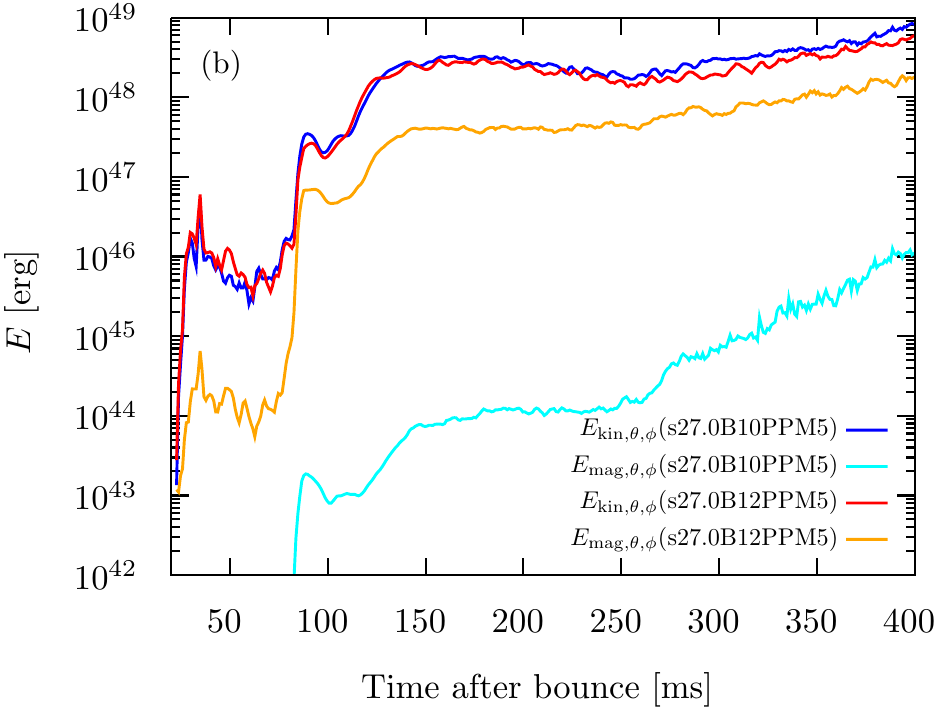}}}
\caption{Time evolution of non-radial components of the kinetic and magnetic turbulent energy
in the gain region for the fiducial progenitor model (s27.0).}
\label{fig9}
\end{center}
\end{figure*}

\begin{figure}
\begin{center}
\scalebox{0.9}{{\includegraphics{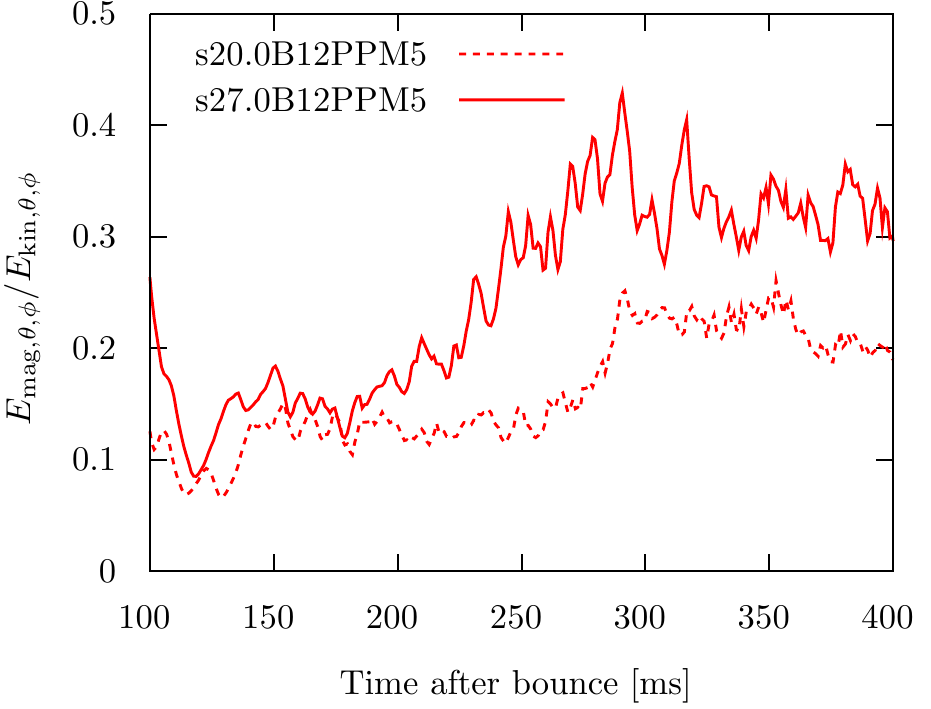}}}
\caption{Time evolution of the ratio between the kinetic and magnetic
turbulent energy in the gain region for model s27.0B12PPM5 (solid line) and s20.0B12PPM5 (dashed line).}
\label{fig10}
\end{center}
\end{figure}

Here, a simple question is how the magnetic field strength could impact the neutrino-heating efficiency to trigger the explosion in our models. Therefore, we compare the time evolution of (a) neutrino luminosity and (b) neutrino mean energy between the s27.0 series model in Fig.~\ref{fig7}. The solid, dashed and dash-dotted lines represent $\nu_e$, $\bar{\nu}_e$ and $\nu_X$, respectively. Here, $\nu_X$ represents the heavy-lepton neutrinos. Blue, red and green lines are the case for models s27.0B10PPM5, s27.0B12PPM5 and s27.0B12PLM2, respectively. As is observed in our previous 2D modelings \citep[see Fig.~3 in][]{ Matsumoto20}, the initial magnetic fields of the progenitor have little impact on the luminosities and mean energies of neutrinos because such properties mainly depend on the mass accretion rate to the PNS. In the model s20.0 series, the magnetic field strength of the progenitor also does not make a large difference of neutrino properties especially before the shock revival.

The gain mass is directly related to the convective motion that delays the mass accretion in the radial direction. The more active the convective motion, the larger the gain mass. As is already mentioned in the introduction, the non-radial flow of the convection is the important key for the enhancement of the efficiency of the neutrino heating. In Fig.~\ref{fig8}, the evolution of the gain mass in $27$ $M_{\odot}$ progenitor models are shown. Comparing different field strength models, the gain mass of the strong field model (s27.0B12PPM5, red line) is smaller than that of the weak field model (s27.0B10PPM5, blue line) after the onset of the shock revival ($t_{\rm pb}=250$\,ms). This result is consistent with that in our previous 2D simulations \citep[see Fig.~2d in][]{Matsumoto20}. The magnetic tension relatively prevents the development of the convection in the strong field model. This results in the difference of the gain mass between different field strength models that is also observed in $20$ $M_{\odot}$ progenitor models.

In order to quantitatively estimate the energy conversion of neutrino heating into the turbulence between different field strength models, the time evolution of non-radial components of the turbulent kinetic and magnetic energies in the gain region for $27$ $M_{\odot}$ progenitor models are shown in Fig.~\ref{fig9}. The non-radial components of the turbulent energy are defined by the sum of the polar ($\theta$-) and azimuthal ($\phi$-) components of the energy. In panel (a), the sum of the turbulent kinetic and magnetic energies for the strong (red line) and weak (blue line) field models are plotted. The evolution of the two lines almost overlaps. This indicates that the efficiency for the conversion of the neutrino heating into the turbulent energy including magnetic fields is independent of the strength of the initial magnetic field of the progenitor. It is reasonable because the magnetic field has little impact on the properties of the neutrinos such as the luminosity and mean energy during the core collapse as shown in Fig.~\ref{fig7}. We, here, stress that the independence of the efficiency for the conversion between neutrino-heating and turbulence energy including the magnetic fields is different from the 3D modeling of \citet{Muller20b}. Although they compare the purely HD and MHD calculations in slowly rotating progenitor of a $15$ $M_{\odot}$ star, the conversion efficiency is enhanced in the MHD model compared to that in HD model.

In panel (b) of Fig.~\ref{fig9}, the turbulent kinetic and magnetic energies in models s27.0B10PPM5 and s27.0B12PPM5 are separately depicted. The blue and cyan lines represent the non-radial components of the turbulent kinetic and magnetic energy in the model s27.0B10PPM5, respectively while the red and orange lines represent the turbulent kinetic and magnetic energy in the model s27.0B12PPM5, respectively. The turbulent magnetic energy in model s27.0B10PPM5 (cyan line) is much less than the turbulent kinetic energy (blue line) in the same model indicating that the impact of the magnetic field on the explosion in our weak field model (s27.0B10PPM5) is passive. The turbulent kinetic energy in the strong field model (red line) is small compared to that in the weak field model (blue line). This is consistent with the relation of the gain mass between the strong and weak field models shown in Fig.~\ref{fig8}.

\begin{figure*}
\begin{center}
\scalebox{0.24}{{\includegraphics{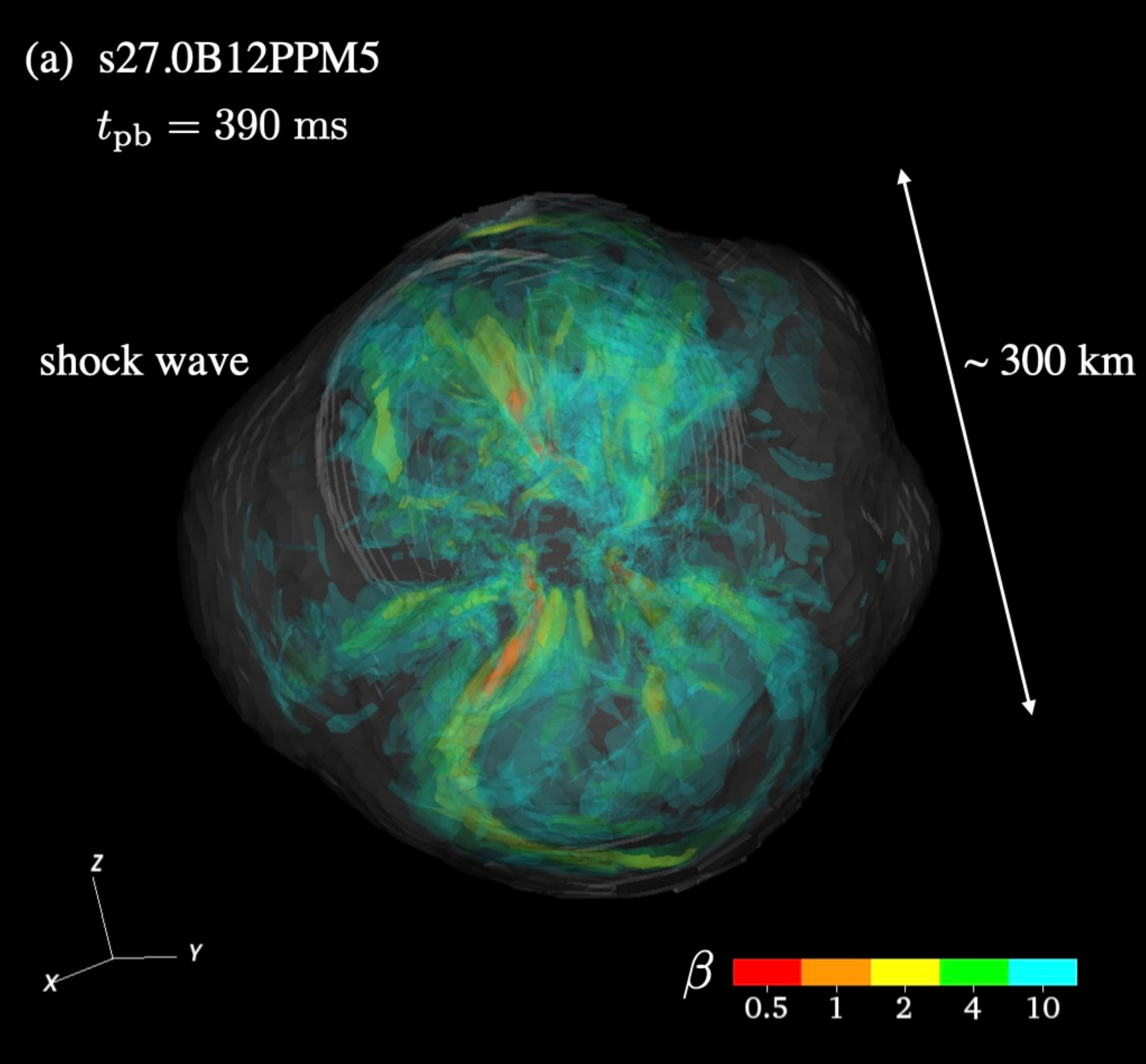}}}
\scalebox{0.24}{{\includegraphics{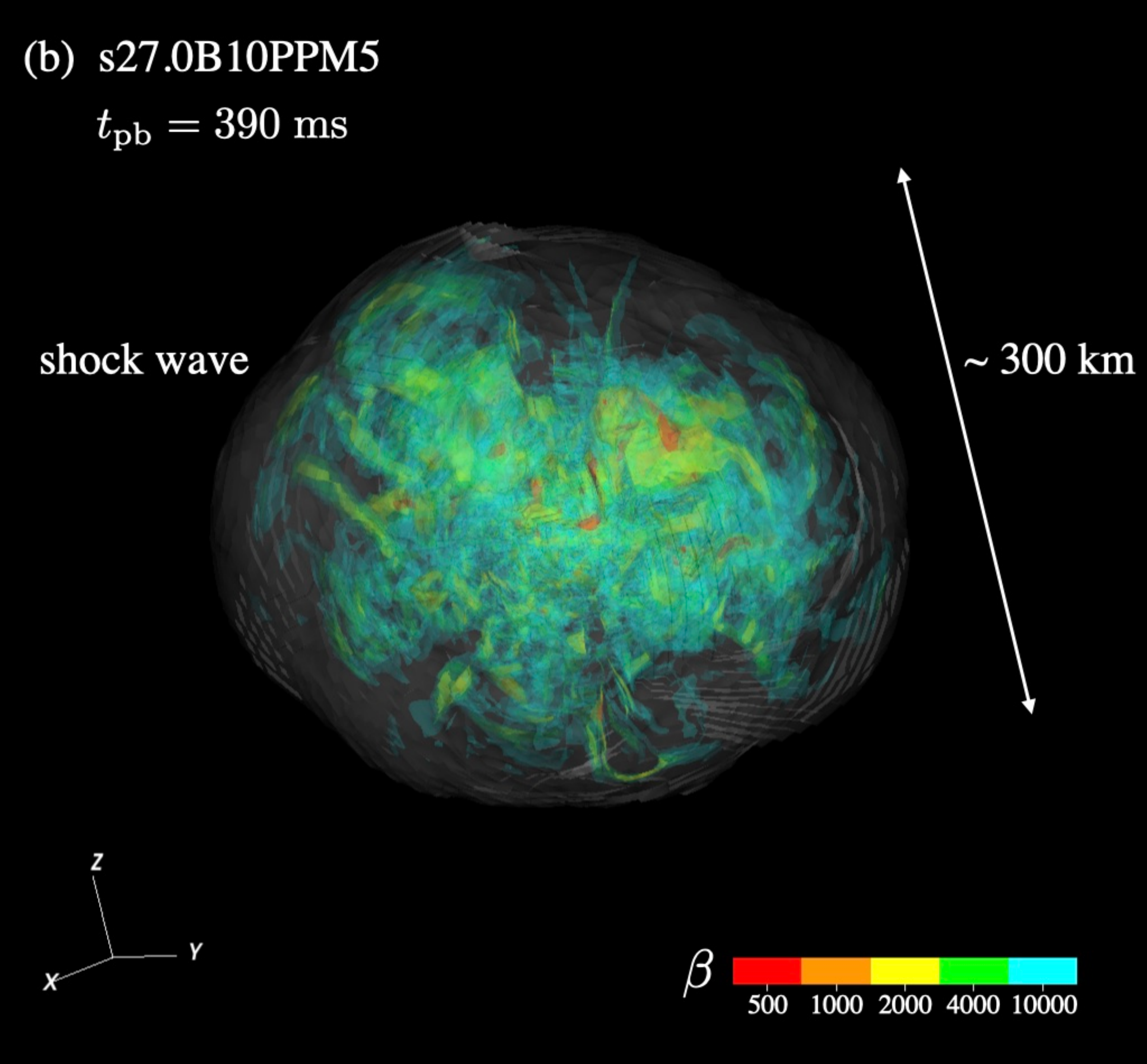}}}
\caption{3D distribution of plasma $\beta$ inside the shock at $t_{\rm pb} = 390$\,ms for model s27.0B12PPM5 (panel a) and s27.0B12PPM5 (panel b). A whitish shell in each panel represents the position of the shock wave.}
\label{fig11}
\end{center}
\end{figure*}

\begin{figure}
\begin{center}
\scalebox{0.4}{{\includegraphics{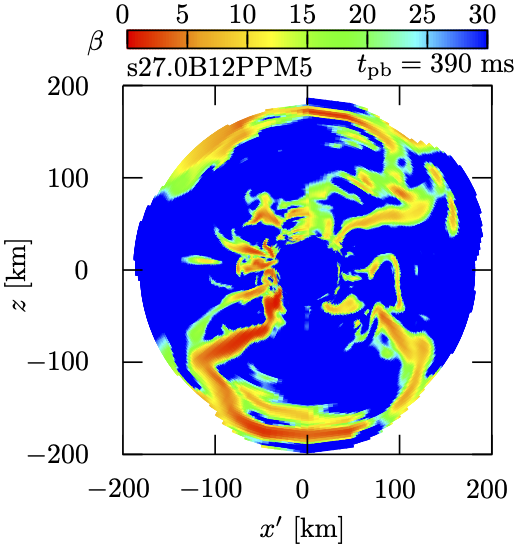}}}
\caption{2D distribution of plasma $\beta$ inside the shock on the cutting plane along the $z$-axis at $t_{\rm pb} = 390$\,ms for the fiducial model (s27.0B12PPM5). Note that $x^{\prime}$ means one axis, whose direction is along $\phi=1.94$ on $x$-$y$ plane and origin matches that of $x$.}
\label{fig12}
\end{center}
\end{figure}

Fig.~\ref{fig10} shows the temporal evolution of the ratio between the turbulent kinetic and magnetic energies in the gain region for strong field models. The red solid and dashed lines correspond to the model s27.0B12PPM5 and s20.0B12PPM5, respectively. After the amplification of the magnetic field due to the convective motion, the energy ratio, $E_{{\rm mag},\theta,\phi}/E_{{\rm kin},\theta,\phi}$, reaches $\sim 0.3$ and $0.2$ in model s27.0B12PPM5 and s20.0B12PPM5, respectively. In the weak field models, it is $\mathcal{O}(10^{-3})$. On the other hand, \citet{Muller20b} reports that the turbulent magnetic energy in the gain layer reaches to $\sim 50$\,\% of the turbulent kinetic energy in the 3D successful explosion model although the initial magnetic field strength is less than that in our weak field models. This is the main driver to explode the $15$ $M_{\odot}$ progenitor that is the failed model in their HD calculation. The growth rate of the turbulent magnetic energy before the runaway shock expansion is obviously different between the MHD model in \citet{Muller20b} and our weak field models. From Fig.~1(c) in \citet{Muller20b}, the exponential growth rate of $E_{B} (\propto \exp(\sigma t))$ between $t=140$\,ms and $t=210$\,ms is roughly estimated to $\sigma = 0.2\,{\rm ms}^{-1}$ while that of $E_{{\rm mag},\theta,\phi}$ between $t=250$\,ms and $t=400$\,ms in the model s27.0B10PPM5 is roughly $\sigma =0.03\,{\rm ms}^{-1}$ as shown in Fig.~\ref{fig9}(b).

Although it is difficult to clarify the origin of the discrepancy of the growth rate of the turbulent magnetic energy because of a complex combination of many factors, one possibility could be the difference of the angular resolution to explain it. The numbers of the grid points in the $\theta$- and $\phi$-direction in our calculations are the half of those in \citep{Muller20b}. In general, the coarser the grid resolution, the larger the numerical diffusion of the magnetic field in MHD calculations. It leads to the slow amplification of the magnetic field. To the contrary, HLLD Reimann solver implemented in our code reduces the numerical diffusion of the magnetic field compared to HLLC Reimann solver used in \citep{Muller20b}. The progenitor dependence on the mass accretion rate may impact on the field amplification. This is because the ram pressure outside the stalled shock that is related to the mass accretion rate decides the radius of it. As explained later (in the section 3.2), the smaller radius of the stalled shock results in the faster growth of the SASI. The difference in the development of the SASI- and/or convection-induced turbulence in progenitors may play an important role in the amplification of the magnetic field. In addition, the formation of the Alfven surface \citep{Guilet11} could contribute to the field amplification in the gain region.

The origin of the fast explosion of the strong field model is expected to be the amplification of the magnetic field in the convective region behind the shock. Fig.~\ref{fig11} shows the 3D distribution of the plasma $\beta$ inside the shock at $t_{\rm pb} = 390$\,ms (soon after the shock revival) in model s27.0B12PPM5 (panel a) and s27.0B10PPM5 (panel b). The plasma $\beta$ is defined by the ratio of the gas pressure to the magnetic pressure and is a good indicator to investigate the dynamics of the fluid. Five and transparent isosurfaces of the plasma $\beta$ are illustrated. The red, orange, yellow, green and cyan represent $\beta=0.5$, $1$, $2$, $4$ and $10$ in panel (a), respectively. Note that the color scale of the plasma $\beta$ in panel (b) is $10^{3}$ times higher than that in panel (a). An outermost whitish and transparent sphere depicts the position of the shock wave. The viewing angle, the spacial scale and the post-bounce time in Fig.~\ref{fig11} are the same as those in Fig.~\ref{fig2}(c).

The value of the plasma $\beta$ behind the shock in the weak field model (panel b) is larger than $10^{4}$. This means that the magnetic field does not have significant impact on the dynamics of the shock expansion at all.

On the other hand, in the strong field model (panel a), the plasma $\beta$ behind the shock surface except the equatorial region ($z \sim 0$) is $\mathcal{O}(10)$. The radial collapse of the initial magnetic field given by the equation (\ref{eq: initial vector potential}) makes the magnetic field around the equatorial region weak. This results in larger value of the plasma $\beta$ there. The relatively small plasma beta region (green and yellow) traces the large bubble structures induced by the neutrino-driven convection. The size of green loops of small plasma $\beta$ as shown in Fig.~\ref{fig11}(a) obviously matches the typical length scale of hot large bubbles of the entropy per baryon in Fig.~\ref{fig2}(c). The accumulation of the magnetic field lines on the interface of the large bubble due to the convective motion leads to the amplification of the magnetic field. Moreover, the strongly magnetized fluid (red and orange in plasma $\beta$ distribution) is observed at the down flow region between the hot bubbles. Since the down flow converges, the magnetic field is highly amplified there.

To show these clearly, 2D spatial distribution of the plasma $\beta$ on the cutting plane along the $z$-axis of Fig.~\ref{fig11}(a) is depicted in Fig.~\ref{fig12}. Note that the color scale of this figure is slightly different from that in Fig.~\ref{fig11}(a). The horizontal axis, $x^{\prime}$, represents one axis, whose direction is along $\phi=1.94$ on $x$-$y$ plane and origin matches that of $x$. One can observe the amplification of the magnetic field in the form of large ($\sim 100$\,km) bubbles at the north ($z>0$) and south ($z$>0) region in Fig.\ref{fig12}. Since the magnetic field does not make a large impact on the shock expansion in the weak field model, the additional support of the magnetic pressure and/or tension on large hot bubbles just behind the shock is the origin of the fast explosion in the strong field model.

\subsection{Impact of spatial accuracy of simulations} \label{dependence of spatial accuracy}
As explained in the previous section, the neutrino-driven explosion occurs in our models. To enhance the neutrino-heating efficiency, the development of the convection is an essentially important process. It is widely known that the development of the convection in numerical simulations generally depends on the spatial accuracy of the calculation. The higher spatial accuracy, the more active the convective motion especially in the small length scale. In this section, we compare the impact of the spatial accuracy on the explosion between strongly magnetized $27$ $M_{\odot}$ progenitor models (s27.0B12PPM5 and s27.0B12PLM2). In our calculations, 5th-order spatial accuracy is obtained in models using PPM5 method while 2nd-order accuracy in the model using PLM2.

As is expected, the higher order accuracy in space is positive for the explosion in our calculations. As shown in Fig.~\ref{fig3}(a), the shock evolution in model s27.0B12PPM5 (red line) is faster than that in model s27.0B12PLM2 (green line). Note that the initial magnetic field strength is the same in both models. In addition, Fig.~\ref{fig5} shows that the diagnostic explosion energy of s27.0B12PPM5 (red solid line) is large compared to that of s27.0B12PLM2 (green solid line).

The difference of the development of the convection crucially affects the explodability. The enhancement of the gain mass in models using PPM5 method (compare red and green lines in Fig.~\ref{fig8}) clearly presents an evidence that the difference of the spatial accuracy of the simulations contributes to the development of the convection. The time evolution of the gain mass in model s27.0B12PLM2 (green line) gradually deviates from that in s27.0B12PPM5 (red line) after around $t_{\rm pb}=200$\,ms. This tendency is closely linked to the shock evolution (see Fig.~\ref{fig3}b). The similar evolution is also observed in Fig.~\ref{fig6}. The timing when the ratio of the advection timescale to the heating timescale exceeds unity in s27.0B12PLM2 (green line) is slightly delayed compared to that in model s27.0B12PPM5 (red line). Therefore, the matter behind the shock in model s27.0B12PPM5 gains the thermal energy more effectively than that in model s27.0B12PPL2. This is the origin of the fast explosion in the model with the higher order accuracy in space.

The spatial accuracy also impacts on the growth of the SASI in our calculations. The oscillation of the shock evolution due to the growth of the SASI is observed in all models after $t_{\rm pb}=250$\,ms as shown in Fig.~\ref{fig3}(b). The amplitude of the oscillation induced by the SASI in model s27.0B12PLM2 (green line) is larger than that of other models. This mechanism can be explained from the size of the shock radius. The shock position in model s27.0B12PLM2 locates inward compared to that in other models at around the onset of the shock revival ($t_{\rm pb} = 250$\,ms). Therefore, the amplitude of the oscillation due to the SASI in model s27.0B12PLM2 becomes large. This is because the SASI grows through the repetition of the advection of vorticity perturbations generated by the deformation of the shock surface and the outward propagation of acoustic waves between the shock and PNS \citep[advective-acoustic cycle,][]{Foglizzo07}. The smaller the shock radius, the shorter the both the advection and propagation time. This leads to the fast grow of the SASI.

The shock surface in model s27.0B12PLM2 expands outwardly due to the SASI compared to that in model s27.0B12PPM5 in a while after the onset of the shock revival (around $t_{\rm pb}=300$\,ms). However, in the late phase, the neutrino-driven convection in the gain region behind the shock wave contributes to shock expansion effectively. Then, the shock expansion in model s27.0B12PPM5 becomes faster than that in model s27.0B12PLM2 after $t_{\rm pb}=350$\,ms.

\begin{figure}
\begin{center}
\scalebox{0.9}{{\includegraphics{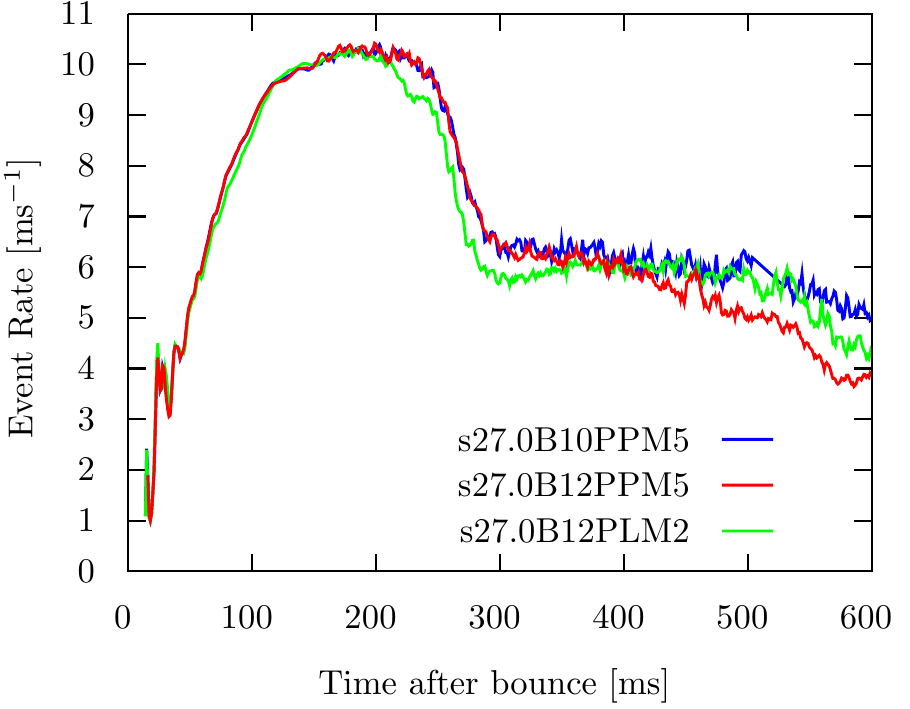}}}
\caption{Event rate of $\bar{\nu}_e$ detected in SK for $27$ $M_{\odot}$ progenitor models. A source distance is set as $10$ kpc. The blue, red and green lines are the case for models s27.0B10PPM5, s27.0B12PPM5 and s27.0B12PLM2, respectively.}
\label{fig13}
\end{center}
\end{figure}

No large difference is observed in the whole evolution of the neutrino luminosity and mean energy focusing on the spatial accuracy of the simulations. The temporal evolution of the neutrino luminosity with different accuracy in space almost overlaps as shown in Fig.~\ref{fig7}(a) (compare red and green lines). On the other hand, in the evolution of the mean energy of neutrinos, all types of green lines in Fig.~\ref{fig7}(b), that is, $\nu_e$, $\bar{\nu}_e$ and $\nu_X$ in model s27.0B12PLM2, show small dips after $t_{\rm pb}=200$\,ms compared to red lines (s27.0B12PPM5) though they do not significantly impact on the dynamics of the explosion. Since the gain mass in model s27.0B12PLM2 is relatively small, the mass accretion rate to the PNS in model s27.0B12PLM2 becomes large compared to that in models using PPM5 method. Therefore, the radius of the neutrino sphere of each type neutrino is expected to become large. The temperature of the neutrinos, that is, the mean energy, compensates and decreases under the condition of the same luminosity. This leads to the small dips in the temporal evolution of the mean energy of neutrinos.

\subsection{Signals of neutrinos and gravitational waves from non-rotating MHD core-collapse model} \label{signals}
\begin{figure}
\begin{center}
\scalebox{0.68}{{\includegraphics{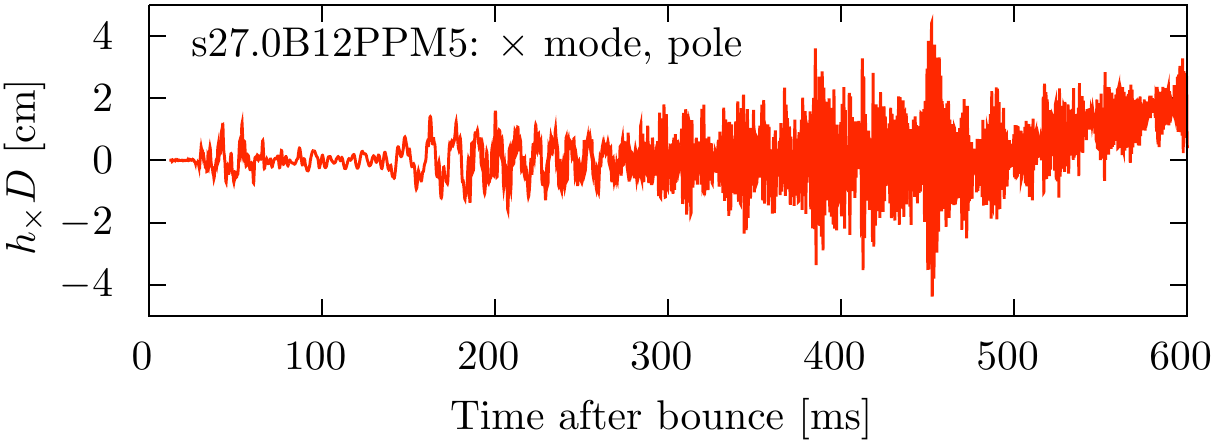}}}
\caption{The gravitational waveform of the $\times$ mode for the fiducial model (s27.0B12PPM5).}
\label{fig14}
\end{center}
\end{figure}

\begin{figure}
\begin{center}
\scalebox{0.35}{{\includegraphics{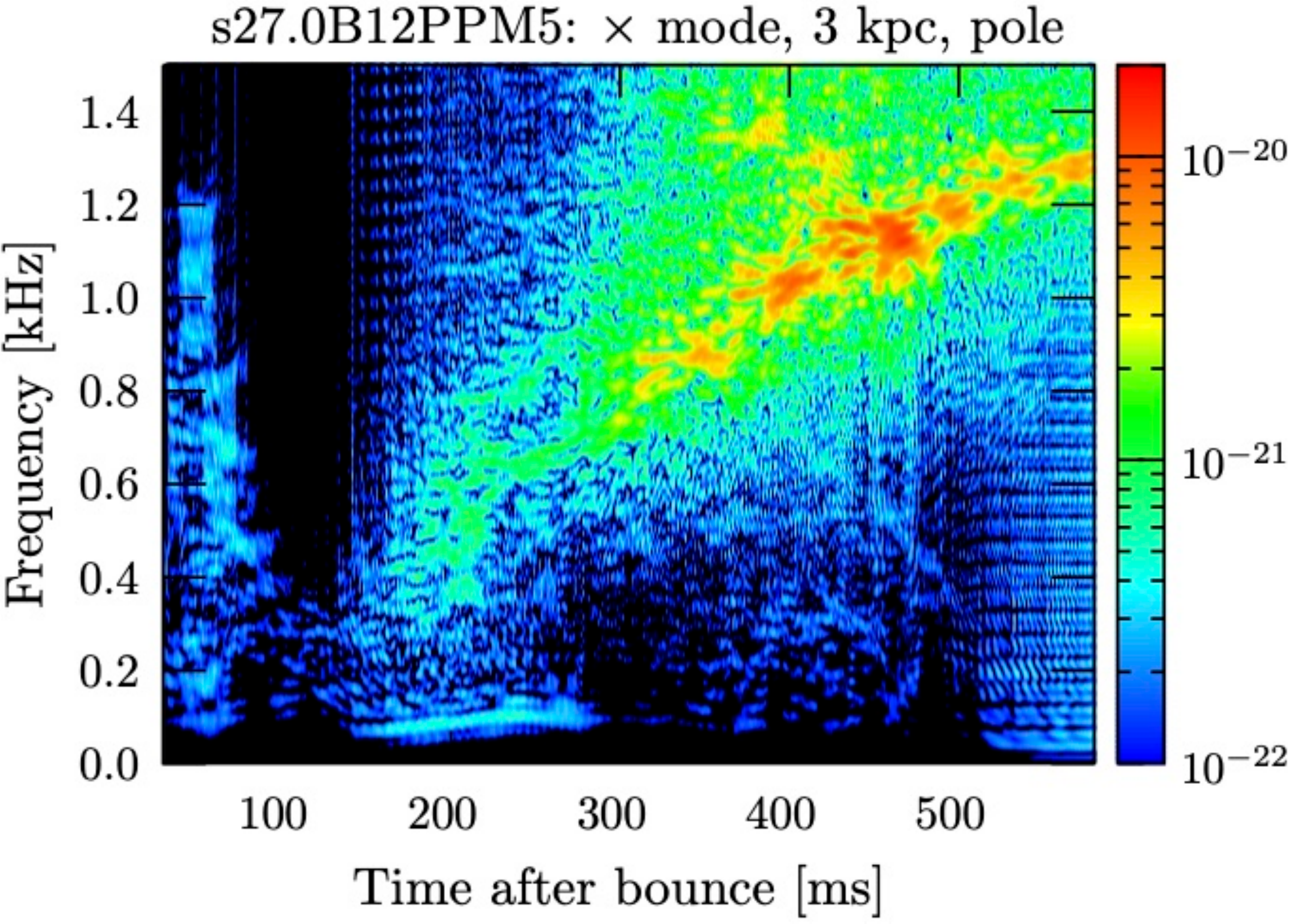}}}
\caption{Spectrogram of the characteristic GW amplitude for model s27.0B12PPM5. The observer is located at a distance of 3\,kpc along the $z$-axis of the source.}
\label{fig15}
\end{center}
\end{figure}

After detecting neutrinos from SN 1987A, the CCSNe are expected to be neutrino emitter candidates. In addition, they are also possible candidates for the detection of GWs. The key information of the physical processes deep inside the SN core would be imprinted into the observational signature of neutrinos and GWs (see \citet{mirrizzi,horiuchi,kotake13,ernazar2020} for a review). In this section, we compute the observational signals of neutrinos and GWs from non-rotating MHD core-collapse in the fiducial progenitor models (s27.0). In this work, we consider the observational signals of only the $\times$ mode of GWs because GW signatures of the $+$ mode is similar to the $\times$ mode in our models.

Fig.~\ref{fig13} shows the temporal evolution of the event rate of $\bar{\nu}_e$ detected in Super-Kamiokande (SK) for $27$ $M_{\odot}$ progenitor models. We calculate the expected detection rates of $\bar{\nu}_e$ following \citet{Takiwaki21} (see equation 30 in their paper for the details). Note that the neutrino detector Hyper-Kamiokande was assumed in \citet{Takiwaki21}, whereas we focus on a currently-working SK (the fiducial volume $32$ kton) in this study. A source distance is set as $10$ kpc in this plot. The blue, red and green lines correspond to models s27.0B10PPM5, s27.0B12PPM5 and s27.0B12PLM2, respectively. Overall evolution between models is not so different. The small deviation of the evolution between models is closely linked to the evolution of the neutrino luminosity and mean energy of the neutrino as shown in Fig.~\ref{fig7}(a) and Fig.~\ref{fig7}(b). It is reasonable because the event rate is estimated using the neutrino luminosity and mean energy.

In not only the neutrino signals but also GW signals, there is no large difference of the model dependence. In Fig.~\ref{fig14}, a representative gravitational waveform of the $\times$ mode for the fiducial model (s27.0B12PPM5) is plotted. An observer is assumed to be located at a distance ($D$) of $3$ kpc along the $z$-axis of the source. This is because it has been reported that the waveform is less sensitive to the viewing angle in the 3D non-rotating models \citep{Andresen17, Takiwaki21}. Actually, the waveforms seen from the pole and the equator are not so different in our models.

\begin{figure}
\begin{center}
\scalebox{0.84}{{\includegraphics{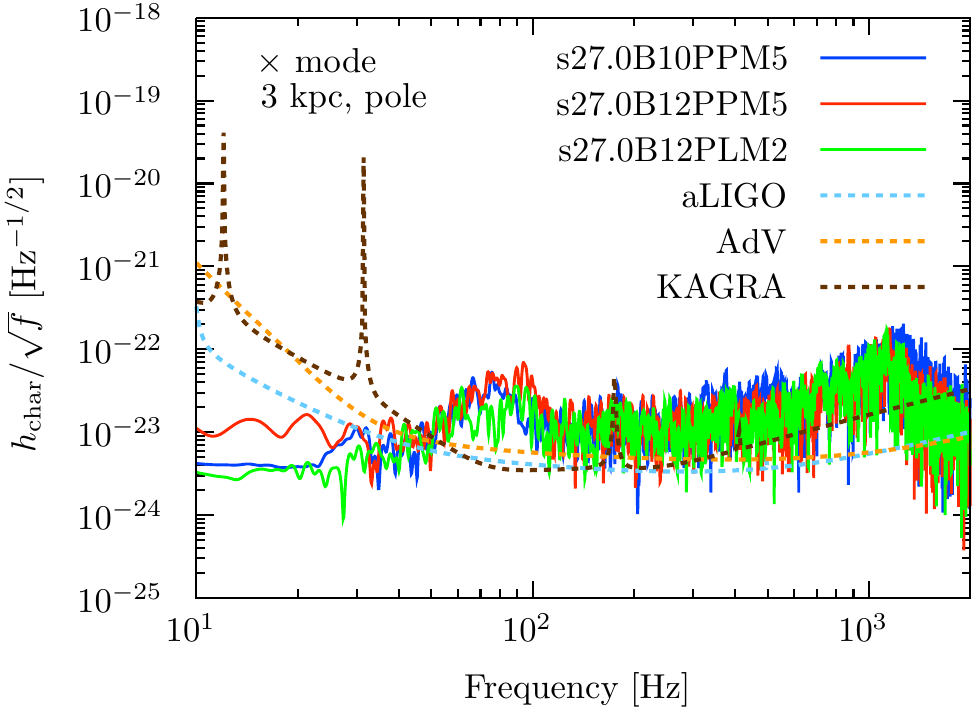}}}
\caption{Characteristic GW spectral amplitudes of $27$ $M_{\odot}$ models shown along the $z$-axis. The observer position is the same as that in Fig.~\ref{fig15}. The sensitivity curves of aLIGO, AdV and KAGRA are also plotted  (the data is extracted from the following URL, \url{https://gwdoc.icrr.u-tokyo.ac.jp/cgi-bin/DocDB/ShowDocument?docid=9537}).}
\label{fig16}
\end{center}
\end{figure}

The GW amplitude in model s27.0B12PPM5 is consistent with that in the same progenitor (s27) model in \citet{Andresen17} (see Figure~1 in their paper). We, here, stress that s27 model in \citet{Andresen17} is purely HD model while the model s27.0B12PPM5 in our work includes the magnetic field. The waveform in model s27.0B12PPM5 indicates well-known features during the core-collapse previously identified in \citet{Mezzacappa20,Radice19,Murphy09}. The oscillations around $t_{\rm pb} = 50$\,ms have their origin in the prompt convection before the stall of the prompt shock. The higher frequency oscillations after the explosion ($t_{\rm pb} > 300$\,ms) is observed. In addition, a monotonic rise of the amplitude due to the global asymmetries of the explosion is also observed after $t_{\rm pb} > 500$\,ms.

The spectrogram of the GW amplitude in model s27.0B12PPM5 is illustrated in Fig.~\ref{fig15}. Following \citet{Murphy09,Takiwaki18,Shibagaki20}, we estimate the characteristic GW strain, $h_{\rm char}$. The lower-frequency ($\sim 100$\,Hz) feature of the GW associated with the SASI is seen before the explosion (150\,ms < $t_{\rm pb} < 300$\,ms). This is consistent with small oscillations of the trajectory of the shock radius (see red line in Fig.~\ref{fig3}b). However, the SASI signal is not vigorous in our model and disappears with runaway shock expansion after $t_{\rm pb} = 300$\,ms as reported in \citet{Andresen17,Andresen19} and \citet{Radice19}. The higher-frequency ($400$\,Hz $< f < 1.2$\,kHz) feature associated with the PNS oscillations, that is, the ramp-up of the frequency due to the shrink of the PNS \citep{Radice19} is also observed. 

To discuss the detectability of the GWs in $27$ $M_{\odot}$ progenitor models, the characteristic GW spectral amplitudes for s27.0 series are plotted in Fig.~\ref{fig16}. Blue, red and green lines have the same meanings as those in Fig.~\ref{fig13}. The detector sensitivity curves of the advanced LIGO (aLIGO), advanced VIRGO (AdV) and KAGRA are also drawn by light blue, orange and brawn dashed lines, respectively. As reported in \citet{Andresen17} and \citet{Takiwaki21}, it is difficult for second-generation instruments like aLIGO, AdV and KAGRA to detect the GW in 3D non-rotating models from a distance of $\sim 10$ kpc. However, as shown in Fig.~\ref{fig16}, the GW signals of magnetized CCSNe from a distance of $3$ kpc or less are detectable.

\begin{figure}
\begin{center}
\scalebox{0.22}{{\includegraphics{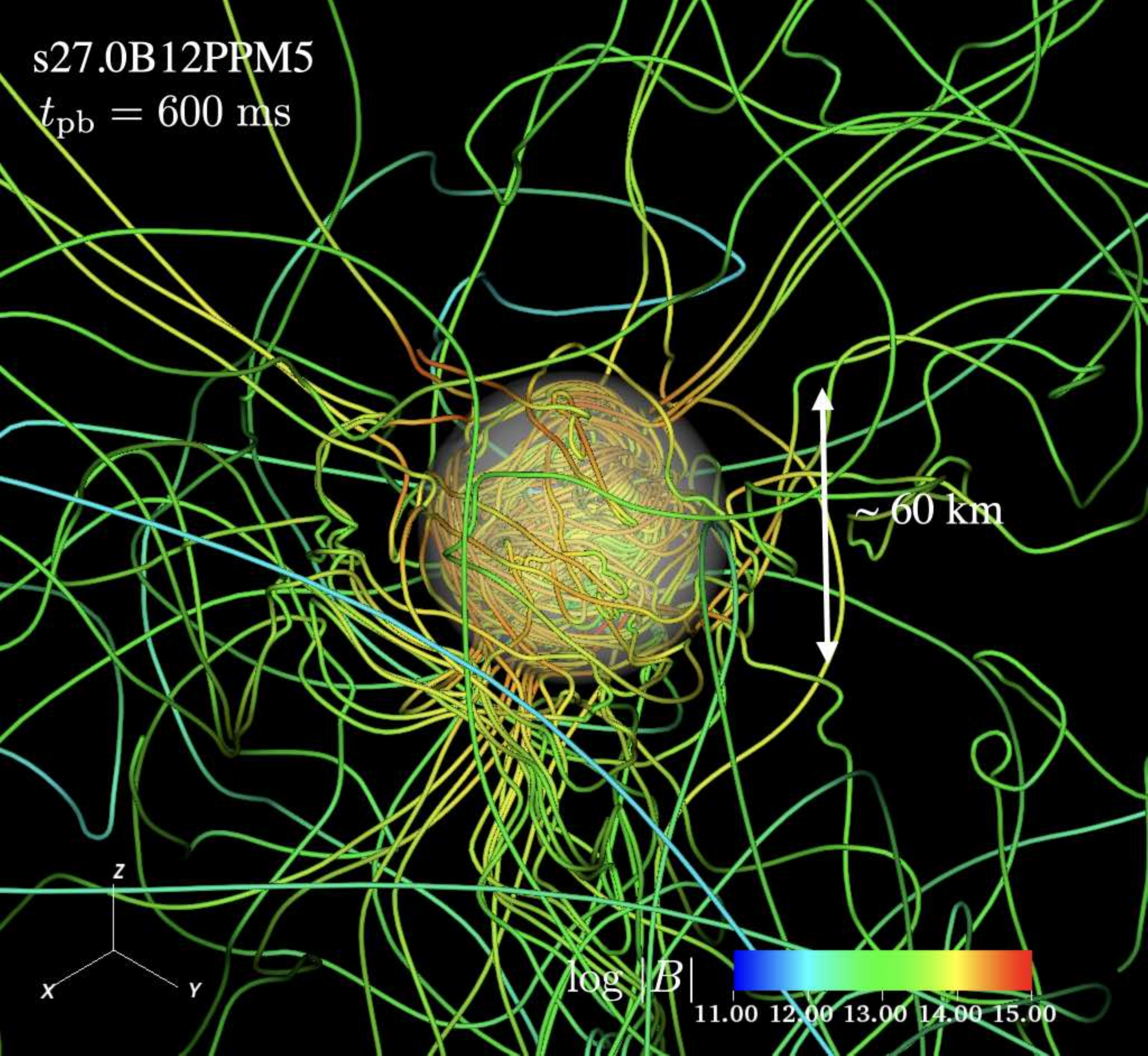}}}
\caption{Magnetic field lines of the PNS at $t_{\rm pb} = 600$\,ms in the model s27.0B12PPM5. The color of magnetic field lines indicates the magnetic field strength in logarithmic scale. A whitish and transparent sphere at the centre represents the iso-density surface of $10^{11}$\,g\,cm$^{-3}$ that is defined as the PNS surface.}
\label{fig17}
\end{center}
\end{figure}

\section{Magnetic field of proto-neutron star} \label{B-field of PNS}
In this section, we shed new light on the origin of the magnetic field of a PNS based on our numerical results. The magnetic field is accumulated and amplified in the convectively stable region beneath the non-rotating PNS surface in all our models. The details of this phenomenon are explained in what follows. 

\begin{figure}
\begin{center}
\scalebox{0.35}{{\includegraphics{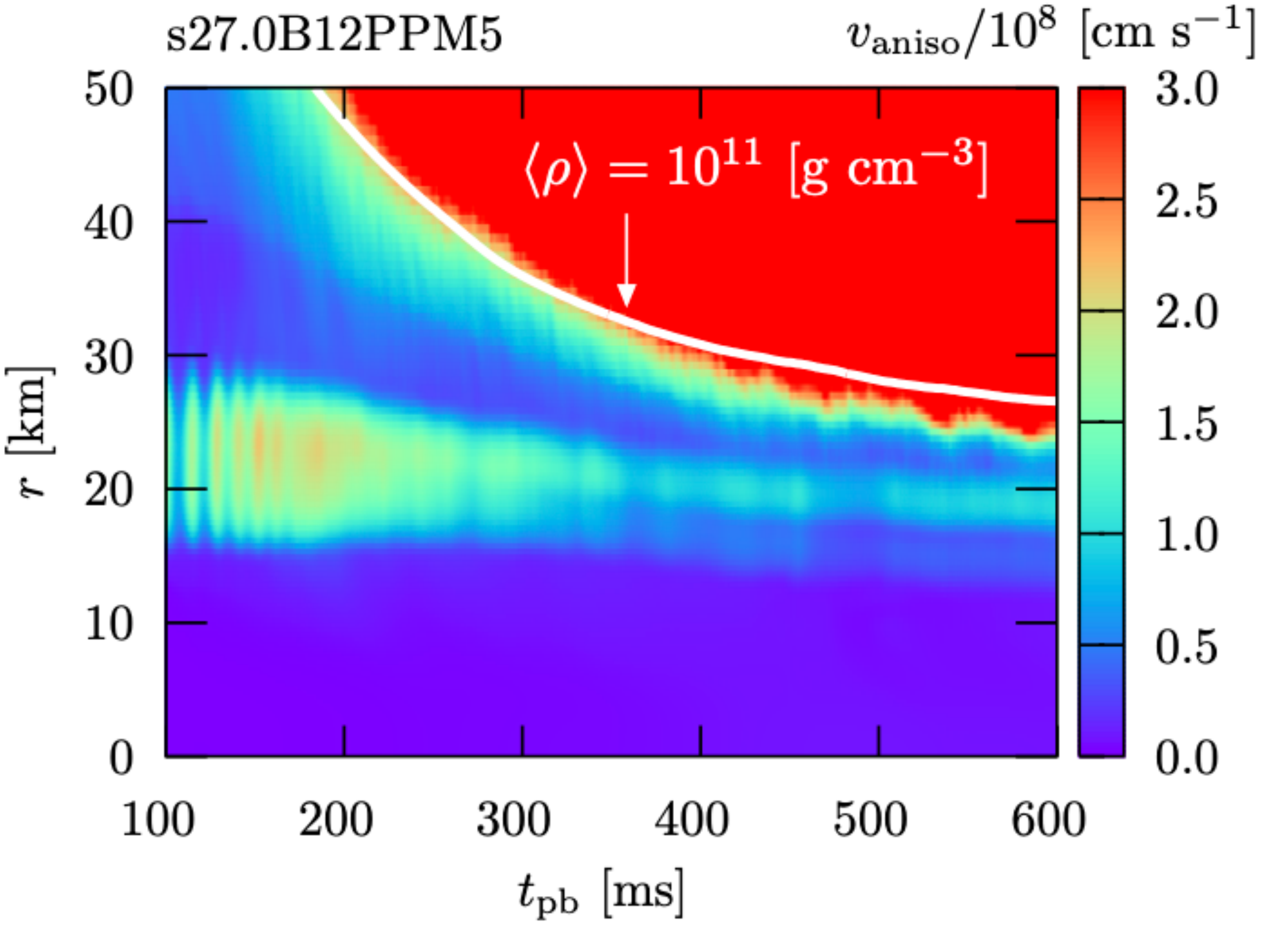}}}
\caption{Time evolution of anisotropic velocity, $v_{\rm aniso}$, defined by equation (\ref{v_aniso}) for model s27.0B12PPM5. The trajectory of the angle average of the density of $10^{11}$\,g\,cm$^{-3}$ is drawn by a white line.}
\label{fig18}
\end{center}
\end{figure}

Fig.~\ref{fig17} shows the 3D configuration of the magnetic field around the PNS at the final phase of the calculation ($t_{\rm pb}=600$\,ms) in the fiducial model (s27.0B12PPM5). The color of magnetic field lines means the magnetic field strength on a logarithmic scale. A whitish and transparent sphere indicates the PNR radius defined by the iso-density surface of $10^{11}$\,g\,cm$^{-3}$. The spatial length scale is represented by a white two-headed arrow that is parallel to the $z$-axis. A magnetar-class magnetic field ($B > 10^{14}$\,G) is observed around the PNS. Since the magnetized core collapses after the start of the simulation and the initial magnetic field that is given by equation (\ref{eq: initial vector potential}) is uniform ($B=B_0$) inside the core ($r < 10^{3}$\,km), by simply considering the magnetic flux conservation, the magnetic field strength around the PNS after the collapse is estimated as follows:
\begin{eqnarray}
B_{\rm PNS} \sim 10^{15}\,{\rm G} \; \Biggl ( \frac{B_0}{10^{12}\,\rm{G}} \Biggr )
\Biggl ( \frac{30\,{\rm km}}{r_{\rm PNS}} \Biggr )^2 \;, \label{eq: estimated B_PNS}
\end{eqnarray}
where $r_{\rm PNS}$ is the radius of the PNS. Therefore, the generation of a magnetar-class magnetic field is reasonable in the fiducial model. The neutrino-driven/PNS convection leads to the complicated structures of magnetic field lines. They are bent because of the convective motion of the fluid and the azimuthal component of the magnetic field is generated although the initial magnetic field is purely poloidal.

\begin{figure*}
\begin{center}
\scalebox{0.23}{{\includegraphics{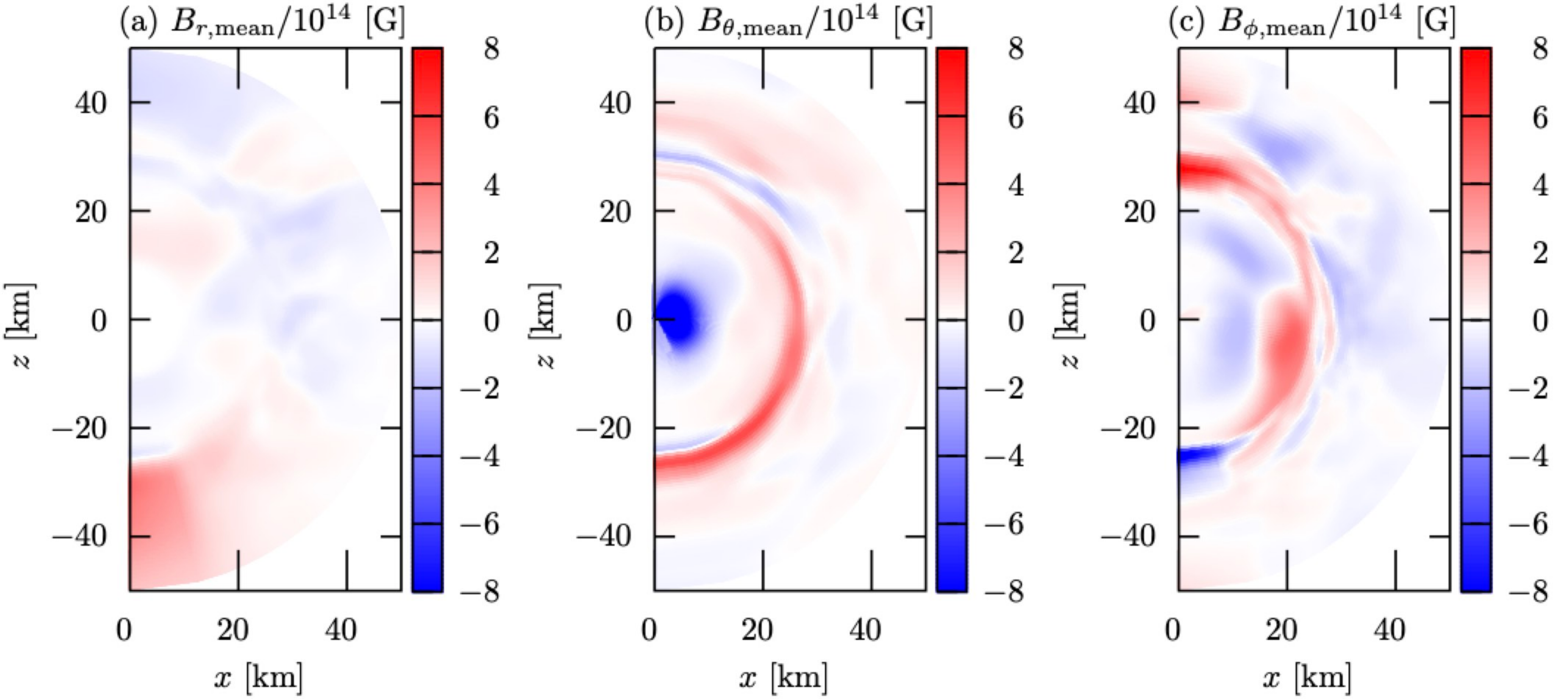}}}
\caption{Meridional distributions of (a) $B_{r,{\rm mean}}$, (b) $B_{\theta,{\rm mean}}$ and (c) $B_{\phi,{\rm mean}}$ defined by equation (\ref{mean magnetic components}) at $t_{\rm pb} = 600$\,ms in the fiducial model (s27.0B12PPM5).}
\label{fig19}
\end{center}
\end{figure*}

\begin{figure}
\begin{center}
\scalebox{0.9}{{\includegraphics{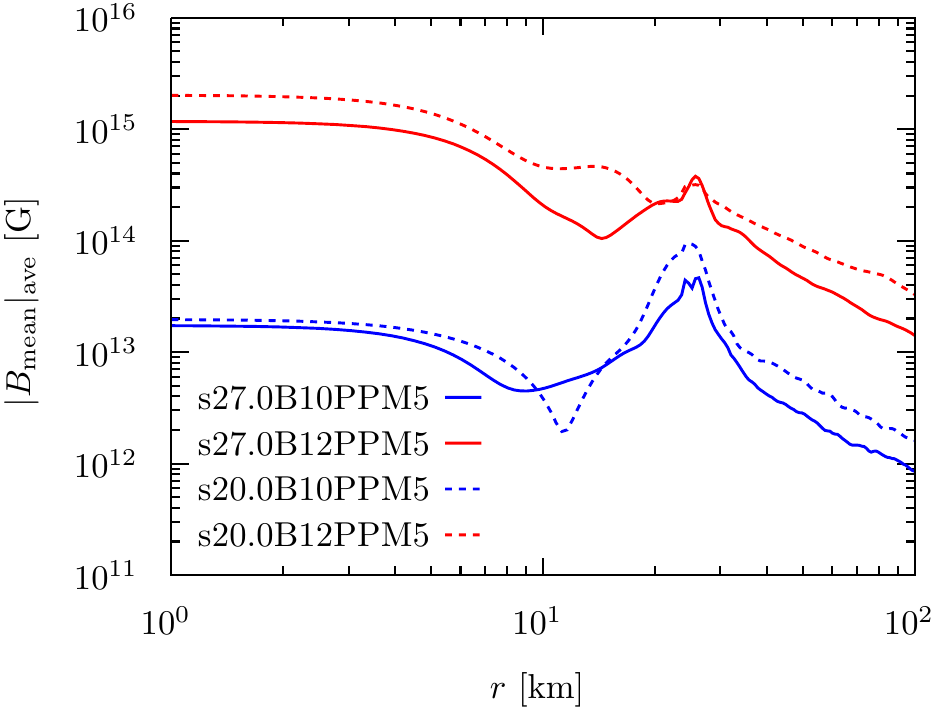}}}
\caption{Radial distributions of the polar angle averaged field strength, which consists of mean magnetic components, $|B_{\rm mean}|_{\rm ave}$, defined by equation (\ref{mean field strength}) at $t_{\rm pb} = 600$\,ms for s27.0 series (solid lines) and $t_{\rm pb} = 500$\,ms for s20.0 series (dashed lines). Red and blue lines correspond to the strong ($B_0 = 10^{12}$\,G) and weak ($B_0 = 10^{10}$\,G) field models, respectively, in each progenitor model.}
\label{fig20}
\end{center}
\end{figure}

In order to measure the strength of convective activity, we compute an anisotropic velocity. Following \citet{Takiwaki12}, it is defined as
\begin{equation}
 v_{\rm{aniso}}=\sqrt{\frac{\langle\rho \lbrack (v_r-\langle v_r \rangle)^2+v_\theta^2+v_\phi^2 \rbrack \rangle}{\langle\rho\rangle}}\; , \label{v_aniso}
\end{equation}
where $\langle v_r \rangle$ and $\langle \rho \rangle$ are the angle average of the radial component of the velocity and the density, respectively.
Here, we define the angle average of variable A as
\begin{equation}
\langle A \rangle = \frac{\int A \sin \theta {\rm d} \theta {\rm d} \phi}{4 \pi} \; .
\end{equation}
The greater deviation in the radial motions ($v_r - \langle v_r \rangle$) and/or larger non-radial ($v_\theta,~v_{\phi}$) motions induces higher anisotropy. The temporal evolution of convective activities ($v_{\rm aniso}$) for model s27.0B12PPM5 is illustrated in Fig.~\ref{fig18}. The evolution of the angle average of the density of $10^{11}$ g cm$^{-3}$ is drawn by a white line that indicates the trajectory of the PNS radius.

The PNS convection due to the negative gradient of the lepton fraction develops between $r=15$\,km and $r=30$\,km after about $t_{\rm pb}=100$\,ms. On the other hand, the site of the development of the neutrino-driven convection, whose origin is the negative gradient of the entropy, is just behind the stalled shock (between $r=100$\,km and $r=150$\,km, see Fig.~\ref{fig3}b). The anisotropic velocity induced by the neutrino-driven convection is $\mathcal{O}(10^{9})$ cm s$^{-1}$ and represented by red color in Fig.~\ref{fig18}. Note that the value of $v_{\rm aniso}$ in the red region is saturated. The white line, that is, the position of the PNS radius, approximately matches the boundary of the red region that penetrates inwards as time passes. At the final phase of the fiducial run (s27.0B12PPM5), the PNS radius shrinks to $\sim 27$\,km. The fluid velocity is highly anisotropic outside the PNS while the fluid beneath the PNS surface ($r \sim 22$\,km) is convectively stable and $v_{\rm aniso}$ is small as shown in Fig.~\ref{fig18}. The PNS convection is located inside this convectively stable region.

To investigate the impact of the convection on the magnetic field strength around the PNS, meridional distributions of mean components of the magnetic field in the azimuthal direction at $t_{\rm pb} = 600$\,ms in the fiducial model are shown in Fig.~\ref{fig19}. The mean magnetic component is given by the longitudinal average as follows:
\begin{eqnarray}
B_{s,{\rm mean}}(r,\theta) = \frac{\int_0^{2 \pi} B_s(r,\theta,\phi) {\rm d} \phi}{2 \pi} \;, \label{mean magnetic components}
\end{eqnarray}
where $s$ stands in the direction along the spherical coordinates ($r,\theta,\phi$). Panel (a), (b) and (c) corresponds to $B_{r,{\rm mean}}$, $B_{\theta,{\rm mean}}$ and $B_{\phi,{\rm mean}}$, respectively. A red (blue) tone in each panel expresses a positive (negative) value of the field strength. In the meridional distributions of mean $\theta$-~ (panel b) and $\phi$-component (panel c) of the magnetic field, a belt where the field strength is larger than that in the surrounding region is observed between $r=20$\,km and $r=30$\,km. The position of the belt overlaps with the convectively stable region beneath the PNS surface as shown in Fig.~\ref{fig18}. The turbulent magnetic diffusion in the PNS convection contributes to the accumulation of the magnetic field in the convectively stable region. In addition, the magnetic field that is advected by the inward flow would secondarily contribute to the surface magnetic fields of the PNS.

The accumulation and amplification of the magnetic field in the convectively stable region beneath the PNS surface is obtained in all our models and is highlighted in Fig.~\ref{fig20}. The radial distributions of the polar angle averaged field strength, $|B_{\rm mean}|_{\rm ave}$, which consists of mean magnetic components, for both $27$ $M_{\odot}$ (solid lines) and $20$ $M_{\odot}$ (dashed lines) progenitor models at the final phase of the calculation are plotted. They are time averages at the centre of $t_{\rm pb} = 600$\,ms for s27.0 series and $t_{\rm pb} = 500$\,ms for s20.0 series. The red and blue lines represent the strong ($B_0 = 10^{12}$\,G) and weak ($B_0 = 10^{10}$\,G) field models, respectively, in each progenitor model. We define $|B_{\rm mean}|_{\rm ave}$ as follows:
\begin{eqnarray}
|B_{\rm mean}|_{\rm ave} = \frac{\int_0^{\pi} \sqrt{B_{r,{\rm mean}}^2 + B_{\theta,{\rm mean}}^2 + B_{\phi,{\rm mean}}^2} {\rm sin}\theta {\rm d}\theta}{2} \; . \label{mean field strength}
\end{eqnarray}

The convectively stable region beneath the PNS surface is located between $r=20$\,km and $r=30$\,km in all models.
The magnetic field is amplified there and becomes larger than that of the surrounding medium. In the strong field models, the magnetar-class magnetic field is organized in the convectively stable shell in the vicinity of the PNS surface. Amazingly, even in the weak field models, the magnetic field strength in the convectively stable shell reaches to $\mathcal{O}(10^{14})$\,G far beyond the anticipated value through the simple magnetic flux conservation during the core-collapse (equation \ref{eq: estimated B_PNS}). 

\begin{figure}
\begin{center}
\scalebox{0.9}{{\includegraphics{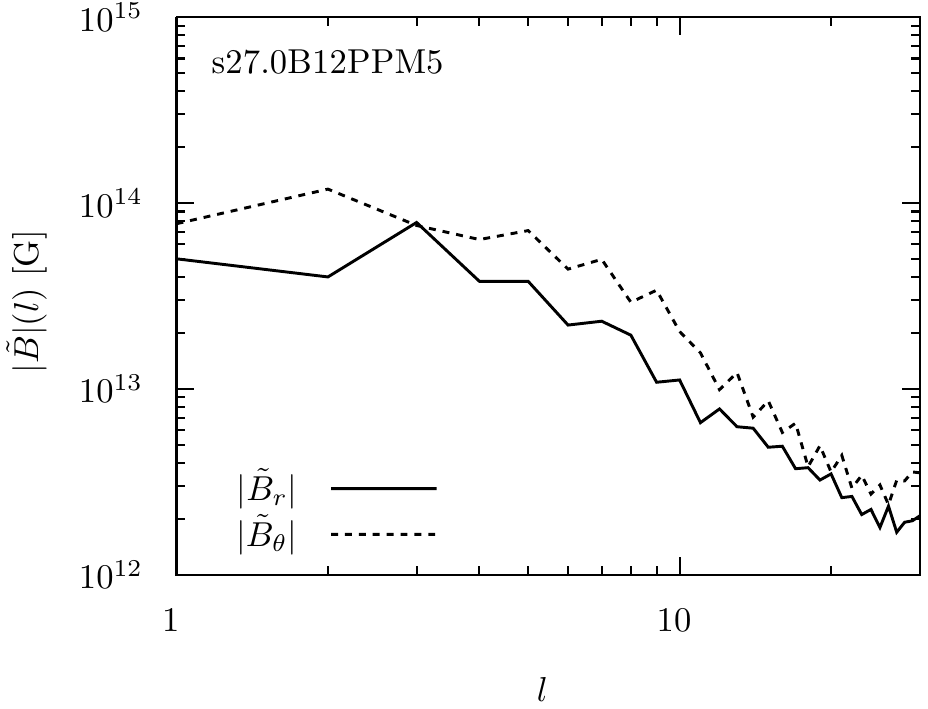}}}
\caption{Spectra of poloidal components of the magnetic field at the convectively stable and strongly magnetized shell ($r=22$\,km) given by equations (\ref{eq: spectrum for Br}) and (\ref{eq: spectrum for B_theta}) in the final phase of the calculation ($t_{\rm pb} = 600$\,ms) for the model s27.0B12PPM5.}
\label{fig21}
\end{center}
\end{figure}

Finally, we compute the spectrum of poloidal components of the magnetic field at the convectively stable and strongly magnetized shell for the fiducial model (s27.0B12PPM5). Plotted in Fig.~\ref{fig21} are
\begin{eqnarray}
\tilde{B}_{r}(l)=\sqrt{\sum_{m=-l}^{l} \biggl | \int_{\Omega} Y_{lm}^{*}(\theta,\phi)
B_{r}(r,\theta,\phi)d\Omega \biggr |^2} \; \label{eq: spectrum for Br}
\end{eqnarray}
and
\begin{eqnarray}
\tilde{B}_{\theta}(l)=\sqrt{\sum_{m=-l}^{l} \biggl | \int_{\Omega} Y_{lm}^{*}(\theta,\phi)
B_{\theta}(r,\theta,\phi)d\Omega \biggr |^2} \;, \label{eq: spectrum for B_theta}
\end{eqnarray}
where $Y_{lm}^{*}$ is the complex conjugate of the spherical harmonics of degree $l$ and $m$, and $\Omega$ is a solid angle. The spectra of poloidal components are evaluated at a fixed radius ($r=22$\,km) in the convectively stable region and at the final phase of the calculation ($t_{\rm pb} = 600$\,ms). The solid and dotted lines represent $\tilde{B}_{r}(l)$ and $\tilde{B}_{\theta}(l)$, respectively. From the complicated structure of the magnetic field lines around the PNS as shown in Fig.~\ref{fig17}, one can apparently anticipate the dominance of the higher order modes in the spectrum. However, in the spectra of both $r$-~ and $\theta$-component, the large scale modes become dominant compared to the small scale modes. In all our models, the large scale magnetic field is dominant in the strongly magnetized shell. Its property of the spectra is similar to that in the PNS convection. The slope of the spectrum is negative as shown in Fig~\ref{fig21}. On the other hand, the slope of the spectrum of the magnetic field in the neutrino-driven convection is rather flat or gentle. Therefore, the impact of the magnetic field amplification in the small-scale length in the neutrino-driven convection on the formation of the strongly magnetized shell is expected to be small. Even though the plenty of matter still fills around the PNS at the final calculation time in our runs, the large scale strong magnetic field in the convectively stable shell beneath the non-rotating PNS radius may be the origin of the magnetic field of the magnetar if the convectively stable region remains until the formation of a crust on the PNS surface.

\section{Summary} \label{summary}
In this paper, we have studied the impact of the magnetic field on the postbounce SN dynamics of non-rotating stellar cores through 3D radiation MHD
simulations with spectral neutrino transport. Initially, $20$ or $27$ $M_{\odot}$ pre-supernova progenitors are threaded by only the poloidal component of the magnetic field. We have considered strong ($B_0=10^{12}$\,G) or weak ($B_0=10^{10}$\,G) field strength for the initial condition in each progenitor model. Focusing on the convective activities induced by neutrino heating, we have compared the explodability between the strong and weak field models.

The highlight in this work is the fast and energetic explosion of the strongly magnetized models though the diagnostic expansion energy is $\mathcal{O}(10^{50})$\,erg. This tendency is observed in both $20$ and $27$ $M_{\odot}$ progenitor models. The efficiency for the conversion of the neutrino heating into the turbulent energy including magnetic fields in the gain region is not significantly different between the strong and weak field models. This results from the independence of the neutrino luminosity and mean energy from the magnetic field strength of the core in this work. They basically depend on the mass accretion rate. The amplification of the magnetic field due to the neutrino-driven convection has large difference between models. The magnetic effect on the dynamics of the system in initially weakly magnetized model is passive. On the other hand, the plasma beta on the surface of large hot bubbles just behind the stalled shock is $\mathcal{O}(10)$ in the strongly magnetized model.The magnetic field is accumulated there because of the convective motion induced by the neutrino heating. Therefore, the amplified magnetic pressure and/or tension can partially contribute to the shock expansion. This is the origin of the fast explosion in the strong field model.

We have also investigated how the difference of the spatial accuracy of the simulation impacts on the explodability in the strong field model for $27$ $M_{\odot}$ progenitor. In our calculations, 5th-order spatial accuracy is obtained in models using PPM5 method while 2nd-order accuracy in the model using PLM2. The higher order accuracy in space contributes to the enhancement of the development of the convection. This leads to the fast and energetic explosion. The spatial accuracy of the simulation does not have a large impact on the neutrino luminosity and mean energy.

In addition to the spatial accuracy of the simulation, the spatial resolution of calculations is an important factor for the property of the explosion. Actually, recent resolution studies for CCSNe without the magnetic field based on the neutrino heating mechanism point out that the higher angular resolution favors the faster explosion reducing the numerical viscosity \citep{Nagakura19b, Melson20}. The higher radial resolution is also expected to contribute to the fast and energetic explosion because of the smaller diffusive transport on the internal energy that keeps the negative entropy gradients \citep{Sasaki21}. The resolution dependence of the MHD modeling of CCSNe based on the neutrino heating is within the scope of our future work.

In our previous 2D modelings of MHD CCSNe \citep{Matsumoto20}, the slightly delayed onset of the shock revival for the strongly magnetized model is obtained using PLM2 method. This may seem to contradict with the fast explosion in the strong field model in our 3D modelings calculated by PPM5. However, the impact of the deference of the dimensionality on the onset of the shock revival is uncertain through only a limited realization of successful MHD explosion. A more extensive realization of the modeling of the MHD CCSN is necessary to draw a robust conclusion to unveil the roles of magnetic fields on the explosion onset in the non-rotating cores.

A new possibility of the origin of the magnetic field of the PNS is proposed based on our results of CCSN simulations for the non-rotating model. The magnetic field is accumulated and amplified to the magnetar level, that is, $\mathcal{O}(10^{14})$\,G, in the convectively stable shell in the vicinity of the PNS surface. Apparently, a much long-term simulation is needed (as  recently done in 2D HD case \citep{Nagakura21} though in different context) to clarify how the global magnetic configuration obtained in this study focusing on the (early) pre-explosion phase would evolve with time to account for the dipolar (or even the subdominant, non-dipolar) configurations of the observed neutron stars/pulsars (e.g., \citealt{enoto19} for a review). To tackle with this question, interesting physics ingredients including the effect of fall-back (e.g., \citealt{ho,vigano,alex,Shigeyama18}), magneto-thermal evolution, not to mention the non-ideal MHD effects (see, \citealt{pons19} for collective references therein) should be carefully treated in numerical computations. This study marks a small, albeit steady, step from the 3D CCSN simulation side toward  understanding the origin of the magnetic fields of pulsars and magnetars.

\section*{Acknowledgements}
We thank Y. Masada, K. Nakamura, Y. Suwa and A. Harada for useful and stimulating discussions.
Numerical computations were carried out on Cray XC50 at the Center for Computational Astrophysics, National Astronomical Observatory of Japan and on Cray XC40 at YITP in Kyoto University. This work was supported by the Keio Institute of Pure and Applied Sciences (KiPAS) project at Keio University, Research Institute of Stellar Explosive Phenomena at Fukuoka University and the associated project (No. 207002), and also by JSPS KAKENHI
Grant Number (JP17K14260, 
JP17H05206, 
JP17K14306, 
JP17H01130, 
JP17H06364, 
JP18H01212, 
JP18K13591, 
JP19K23443, 
JP20K14473, 
JP20K11851, 
JP20H01941, 
JP20H00156 
JP21H01088 
JP21H04488 
and
JP22H01223). 
This research was also supported by MEXT as “Program for Promoting researches on the Supercomputer Fugaku” (Toward a unified view of he universe: from large scale structures to planets, JPMXP1020200109) and JICFuS.

\section*{DATA AVAILABILITY}
The data underlying this article will be shared on reasonable request to the corresponding author.

\bibliographystyle{mnras}
\bibliography{papers} 

\label{lastpage}
\end{document}